\documentclass[10pt,twocolumn]{article}
\usepackage[top=1.8cm, bottom=1.9cm, left=1.8cm, right=1.8cm]{geometry}
\usepackage{times}
\usepackage{booktabs} %
\usepackage{paralist}
\usepackage[small,bf]{caption}
\usepackage[keeplastbox]{flushend}
\usepackage{subfigure}
\usepackage{graphicx}
\usepackage[sort&compress,numbers]{natbib}

\usepackage{tikz}
\usepackage{color,xcolor}
\usepackage{subfigure}
\usepackage{cleveref}
\usepackage{graphicx}
\usepackage{multirow}
\usepackage{graphicx}
\usepackage{amssymb}
\usepackage{amsfonts}
\usepackage{wrapfig}
\usepackage{alltt}
\usepackage{amsmath}
\usepackage{booktabs}
\usepackage{xspace}
\usepackage{natbib}
\usepackage{verbatim}
\usepackage{footnote}
\usepackage{epstopdf}
\usepackage{slashbox}
\usepackage{color, colortbl}

\usepackage[hyphens]{url}
\makeatletter
\def\url@leostyle{%
  \@ifundefined{selectfont}{\def\UrlFont{\small}}%
  {\def\UrlFont{}}%
}
\makeatother
\definecolor{darkgreen}{RGB}{47,109,79}
\definecolor{darkblue}{RGB}{57,79,99}
\usepackage[bookmarks=false, colorlinks=true, plainpages = false,  linkcolor = darkgreen,   citecolor = darkgreen, urlcolor = darkblue, filecolor = blue]{hyperref}

\usepackage[hang,flushmargin]{footmisc}

\renewenvironment{thebibliography}[1]{
  \begin{oldthebibliography}{#1}
    \setlength{\itemsep}{0.1em}
    \setlength{\parskip}{0.1em}
}
{
  \end{oldthebibliography}
}

\renewcommand{\footnoterule}{%
  \kern -4pt
  \hrule width 1in 
  \kern 2pt
}

\renewcommand{\footnotesize}{\fontsize{7}{8}\selectfont}

\usepackage[compact]{titlesec}
\titlespacing*{\section}{0pt}{*3}{3pt} 
\titlespacing{\subsection}{0pt}{*2}{2pt}

\newcommand{\descr}[1]{\vspace{0.1cm}\noindent\textbf{#1}}

\def\mama{\textsc{MaMaDroid}\xspace}
\def\chimp{CHIMP\xspace}
\def\papa{\textsc{AuntieDroid}\xspace}

\makeatletter
\input{t1lmtt.fd}
\@namedef{T1+lmtt}{}
\makeatother
\makeatletter
\newcommand\smallscriptsize{\@setfontsize\scriptsize{5.75}{6.75}}
\makeatother

\usepackage{ifthen}
\newif\ifshort
\shortfalse %

\ifshort
	\newcommand{\isShort}{true}
\else
	\newcommand{\isShort}{false}
\fi

\newcommand{\shortVer}[1]{\ifthenelse{\equal{\isShort}{true}}{{#1}}{}}
\newcommand{\longVer}[1]{\ifthenelse{\equal{\isShort}{false}}{{#1}}{}}

\begin{document}

\pagenumbering{arabic}

\title{\bf A Family of Droids -- Android Malware Detection via Behavioral Modeling: Static vs Dynamic Analysis\thanks{A preliminary version of this paper appears in the Proceedings of 16th Annual Conference on Privacy, Security and Trust (PST 2018). This is the full version.}}
\author{Lucky Onwuzurike$^1$, Mario Almeida$^2$, Enrico Mariconti$^1$,\\Jeremy Blackburn$^3$, Gianluca Stringhini$^1$, Emiliano De Cristofaro$^1$\\[1ex]
\normalsize $^1$University College London $\;\;\;^2$Polytechnic University of Catalonia $\;\;\;^3$University of Alabama at Birmingham}
\date{}

\maketitle

\begin{abstract}
Following the increasing popularity of mobile ecosystems, cybercriminals have increasingly targeted them, designing and distributing malicious apps that steal information or cause harm to the device's owner. 
Aiming to counter them, detection techniques based on either static or dynamic analysis that model Android malware, have been proposed.
While the pros and cons of these analysis techniques are known, they are usually compared in the context of their limitations e.g., static analysis is not able to capture runtime behaviors, full code coverage is usually not achieved during dynamic analysis, etc.
Whereas, in this paper, we analyze the performance of static and dynamic analysis methods in the detection of Android malware and attempt to compare them in terms of their detection performance, using the same modeling approach. 

To this end, we build on \mama, a state-of-the-art detection system that relies on static analysis to create a behavioral model from the sequences of abstracted API calls. Then, aiming to apply the same technique in a dynamic analysis setting, we modify \chimp, a platform recently proposed to crowdsource human inputs for app testing, in order to extract API calls' sequences from the traces produced while executing the app on a \chimp virtual device. We call this system \papa and instantiate it by using both automated (Monkey) and user-generated inputs. We find that combining both static and dynamic analysis yields the best performance, with $F$-measure reaching 0.92. We also show that static analysis is at least as effective as dynamic analysis, depending on how apps are stimulated during execution, and, finally, investigate the reasons for inconsistent misclassifications across methods. 
\end{abstract}

\setcitestyle{numbers,sort&compress}

\vspace*{-0.15cm}
\section{Introduction}
\label{sec:intro}

In today's digital society, individuals rely on smart mobile devices and apps for a plethora of social, productivity, and work activities. %
Inevitably, this makes them valuable targets for cybercriminals, thus, more and more malware are developed every year exclusively targeting mobile operating systems~\cite{androidtrend} and, given its market share~\cite{statistics}, Android in particular.
Compared to malicious software on desktops, malicious apps pose new threats as attackers might be able to, e.g., defeat two-factor authentication of banking systems~\cite{google} or continuously spy on victims through their phone camera or microphone~\cite{independent}. 

As a result, the research community has proposed a number of techniques to detect and block Android malware based on either \emph{static} or \emph{dynamic} analysis. With the former, the code is recovered from the apk, and features are extracted to train machine learning classifiers; with the latter, apps are executed in a controlled environment, usually on an emulator or a virtual device, via a real person or an automatic input generator such as Monkey~\cite{monkey}.
In particular, a few approaches have been recently proposed aiming to improve accuracy of malware detection.
{\em (1) Behavioral Modeling:} Mariconti et al.'s \mama~\cite{mariconti2016mamadroid} builds from static analysis, a behavioral model of malware samples, relying on the {\em sequences} of abstracted API calls;
this yields higher accuracy than state of the art, while also providing higher resilience to API changes and reducing the need to re-train models.
{\em (2) Input Generators:} previous work~\cite{anand2012automated,Machiry2013,carter2016curiousdroid} has introduced input generators that aim to mimic app usage by humans, more effectively than the standard Android pseudorandom input generator (Monkey), thus improving the chances of triggering malicious code during execution.
{\em (3) Hybrid Analysis:} by combining static and dynamic analysis, hybrid analysis has been used to try and get the best of the two worlds, typically, following two possible strategies. One approach is to use static analysis to gather information about the apps under analysis (e.g., intent filters an app listens for, execution paths to specific API calls, etc.) and then ensuring that all execution paths of interest are triggered during the dynamic analysis stage~\cite{carter2016curiousdroid,wong2016intellidroid};
in the other, features extracted using static analysis (e.g., permissions, API calls, etc.) are combined with those from dynamic analysis (e.g., file access, networking events, etc.), and used to train an ensemble machine learning model~\cite{Lindorfer2014,lindorfer2015marvin}. 

\descr{Motivation.} Overall, despite a large body of work proposing various Android malware detection tools, 
the research community's stance on whether to use static or dynamic analysis primarily stems from the systems limitations and the vulnerabilities to possible evasion techniques faced by each approach.
For instance, static analysis methods that extract features from permissions requested by apps often yield high false positive rates, since benign apps may actually need to request permissions classified as dangerous~\cite{Enck2009}, while systems that perform classification based on the frequency of API calls~\cite{Aafer2013DroidAPIMiner} often require constant retraining; moreover, reflection and dynamic code loading can be used to evade static analysis based detection.
On the other hand, the accuracy of dynamic analysis is greatly dependent on whether malicious code is actually triggered during test execution, and in general, dynamic analysis often does not scale.
Nonetheless, we still lack a deep understanding of the advantages and disadvantages of each method in terms of simple detection performance. 

\descr{Roadmap.} In this paper, we aim to fill this research gap by addressing the following research questions: %
\begin{enumerate}
\item Can we extend malware detection techniques based on behavioral modeling (in static analysis, as per \mama~\cite{mariconti2016mamadroid}) to dynamic analysis? %
\item How do different malware analysis methods (i.e., static, dynamic, and hybrid analysis) compare to each other, in terms of detection performance, when the same technique is used to build malware detection models? %
\item Does having humans test apps during dynamic analysis improve malware detection compared to pseudorandom input generators such as Monkey~\cite{monkey}? 
\end{enumerate}

Aiming to answer these questions, we first of all modify \chimp~\cite{mario2018chimp}, a platform allowing to crowdsource human inputs to test Android apps, to support building a behavioral model based malware detection system (as per \mama~\cite{mariconti2016mamadroid}). %
That is, we use the same approach as \mama %
to extract sequences of abstracted API calls from the traces produced while executing the app in a virtual device (instead of the apk). We call this system \papa and instantiate it by using both automated (Monkey) and user-generated inputs.
Then, we evaluate each analysis method, using the same modeling approach (i.e., a behavioral model relying on Markov chains built from the sequences of abstracted API calls), the same features, and the same machine learning classifier.

\descr{Contributions.} %
Overall, we make several contributions. First, we introduce \papa, a virtual device that extends \chimp~\cite{mario2018chimp} and allows for the collection of the method traces (from which features are extracted) produced by an app when executed. Second, we build and evaluate a hybrid system combining behavioral-based static and dynamic analysis features. \longVer{Third, we show that dynamic code loading is prevalent in the wild, which could possibly be used by malware developers to evade static analysis.}
Finally, we compare the different methods, showing that hybrid analysis performs best and that static analysis is at least as effective as dynamic analysis. %

\descr{Paper Organization.} 
The rest of this paper is organized as follows. Next section reviews previous work on Android malware detection.
Then, in Section~\ref{sec:methods}, we describe our modifications to \chimp that enable its use for Android malware detection.
In Section~\ref{sec:setup}, we introduce our experimental setup and the datasets used for evaluation. Next, we present and compare the results achieved by all methods in Sections~\ref{sec:eval} and~\ref{sec:comparison}, respectively.
Finally, the paper concludes in Section~\ref{sec:conclusion}.
(In the Appendix, we also discuss challenges with our dynamic analysis
efforts and the limitations of our virtual device testbed.)

\section{Related Work}
\label{sec:related}
\longVer{We now review previous work on Android malware detection using static, dynamic, and hybrid analysis.}

\subsection{Static Analysis}\label{sec:static}
Android malware detection based on static analysis aims to classify an app as malicious or benign by relying on features extracted from the app's apk, i.e., its source code. 
Techniques presented in~\cite{Enck2009, sanz2013puma, huang2013performance,Sarma2012}
build features from the {\em permissions} requested by the apps, leveraging the fact that malware often tend to request dangerous/unneeded permissions. 
This approach, however, may be prone to false positives, as benign apps may also request dangerous permissions~\cite{Enck2009}.
Moreover, since Android 6.0, the permission model allows users to grant permissions at run-time when they are required, thus some dangerous permissions might never actually be granted (in fact, app developers often request permissions that are never used~\cite{felt2011android}).
Drebin~\cite{arp2014drebin} combines several features extracted %
from the apps' manifest (e.g., intents, hardware components, app components) 
as well as disassembled code (restricted and suspicious API calls, and network addresses) 
to train a classifier. 
Alas, techniques based on decompiled code can be evaded using dynamic code loading, reflection, and the use of native code~\cite{rastogi2014catch,poeplau2014execute}. 

Other tools rely on {\em API calls}. DroidAPIMiner~\cite{Aafer2013DroidAPIMiner} performs classification based on the API calls more frequently used by malware.
However, due to changes in the Android API, as well as the evolution of malware, this requires frequent retraining of the system as new API versions are released and new types of malware are developed. 
Deprecation and/or addition of API calls with new API releases is quite common, and this might prompt malware developers to switch to different API calls. 
Also based on static analysis is \mama~\cite{mariconti2016mamadroid}, which uses behavioral models built from the {\em sequences}, rather than the frequency, of API calls. 
Specifically, it operates by characterizing the transitions between
different API calls, involving the following four stages:
(1) It extracts the call graph of an app, i.e., the control flow graph of the API calls in the apk;
(2) It parses the call graph as sequences of API calls, which are abstracted to one of two modes, to either their ``family'' or package names. In {\tt package} mode, an API call is abstracted to its package name using the list of around 338 
packages from the Android and Google APIs, whereas in {\tt family} mode, to the {\tt google}, {\tt java}, {\tt javax}, {\tt android}, {\tt xml}, {\tt apache}, {\tt junit}, {\tt json}, or {\tt dom} families. Obfuscated and developer specific %
API calls are abstracted to {\tt obfuscated} and {\tt self-defined}, respectively;
(3) Next, it models the sequences of (abstracted) calls as Markov chains, and extracts as features, %
the transition probabilities between states; and finally 
(4) it trains a machine learning classifier geared to label samples as benign or malicious. 

\mama achieves high detection accuracy (up to 0.99 F1-score), and preserves it for longer periods of time compared to~\cite{Aafer2013DroidAPIMiner}, as it builds  models that are more resilient to API changes and malware evolution.
In this paper, for the static analysis part, we build on \mama, re-using the source code publicly available from~\cite{mamadroid_source}, 
to perform and compare malware detection using a behavioral model built from API sequences,
while using both static, dynamic and hybrid analysis (see Section~\ref{sec:setup}). %

\subsection{Dynamic Analysis}
Dynamic analysis based techniques attempt to detect malware by capturing the runtime behavior of an app,
targeting either generic malware behaviors or specific ones.

DroidTrace~\cite{zheng2014droidtrace} uses ptrace (a system call often used by debuggers to control processes) to monitor selected system calls, allowing to run dynamic payloads and classify their behavior as, e.g., file access, network connection, inter-process communication, or privilege escalation. 
Canfora et al.~\cite{Canfora2015} extract features from the sequence of system calls by executing apps on a VM, while Lageman et al.~\cite{lageman2015detecting} model an app's behavior during execution on a VM using both system calls and {\tt logcat} logs. %
CopperDroid~\cite{tam2015copperdroid} uses dynamic analysis to  reconstruct malware behavior by observing executed system calls. %
While CrowDroid~\cite{Burguera2011Crowdroid}, a client running on the device, captures system calls generated by apps and sends them to a central server, which builds a behavioral model of each app.
Whereas,
we build a behavioral model of each app from the sequences of API calls invoked (rather than whether an API call was invoked or not) during execution of the apps.  %

\subsection{Hybrid Analysis}
A few tools combine static and dynamic analysis, e.g., by using the former to analyze an apk and the latter to determine what execution paths to traverse, or by combining features extracted using both static and dynamic analysis. 
Andrubis~\cite{Lindorfer2014} is a malware analysis sandbox %
that \longVer{extracts permissions, services, broadcast receivers, activities, package name, and SDK version from an app's manifest as well as the actual bytecode; it then }%
dynamically builds a behavioral profile \longVer{using static features as well as selected dynamic ones, such as reading/writing to files, sending SMS, making phone calls, use of cryptographic operations, dynamic registration of broadcast receivers, loading {\tt dex} classes/native libraries, etc.} \shortVer{of an app using static (permissions, services, package name etc.) %
and dynamic features (reading/writing to files, sending SMS etc.).} 
Note that although we perform method tracing similar to Andrubis, we use a virtual device and allow humans (and not only Monkey) to test the apps, as the latter perform a random sequence of actions/events which do not necessarily reflect how humans use the apps and may not trigger certain malicious code. 
Also, we build the behavioral profile of the apps from \emph{all} API calls observed in the method traces, rather than selected API calls. 

Marvin~\cite{lindorfer2015marvin} uses features from both static and dynamic analysis to award malice scores (ranging from 0 to 10) to an app and classify as malware, apps with scores greater than 5, while CuriousDroid~\cite{carter2016curiousdroid}, an automated user interface (UI) interaction for Android apps, integrates Andrubis~\cite{Lindorfer2014} as its dynamic module in order to detect malware. 
It decomposes an app's UI on-the-fly and creates a context-based model generating series of interactions that aim to emulate real human interaction. 
IntelliDroid~\cite{wong2016intellidroid} introduces a targeted input generator that integrates with TaintDroid~\cite{Enck2014taintdroid} aiming to track sensitive information flow from a source (e.g., a content provider such as contact list database) to a sink (e.g., network socket). It allows the dynamic analysis tool to specify APIs to target and generates inputs in a precise order that can be used to stimulate the Application Under Analysis (AUA) to observe potential malicious behavior.

Since there are several entry points into an Android app (e.g., via an activity, service, and broadcast), dynamically stimulating an AUA is usually done using tools like Monkey or MonkeyRunner, or humans.
Targeted input generation tools such as CuriousDroid and Intellidroid aim to provide an alternative stimulation of apps that is closer to stimulation by humans and more intelligent than Monkey and MonkeyRunner. %

Finally, we refer the reader seeking more details on the large body of work on Android malware to useful surveys of Android malware families and detection tools in~\cite{amamra2012smartphone, faruki2015android,tam2017evolution}
as well as an assessment of Android analysis techniques in~\cite{reaves2016droid}.

\section{\papa: Behavioral Modeling on a Virtual Device}
\label{sec:methods}

\begin{figure}[t]
 \center
 \includegraphics[width=0.4\textwidth]{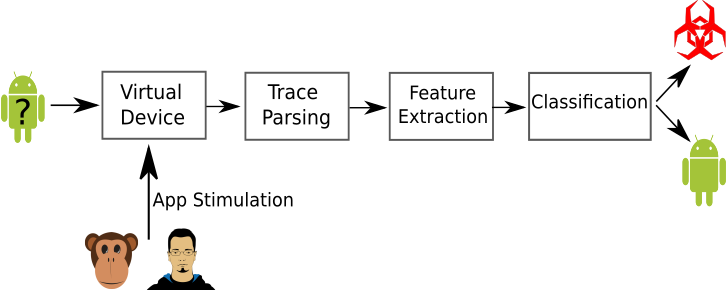}
 \caption{High-level overview of \papa. An apk sample is run in a virtual device, using either Monkey or human. Then, the APIs called during execution are parsed and used for feature extraction. Finally, the app is classified as either benign or malicious.} %
 \label{fig:diagram}

\end{figure}

We now present \papa, a system performing Android malware detection based on behavioral models extracted via dynamic analysis. 
Our main objective is to compare its performance to its static analysis counterpart, i.e., \mama~\cite{mariconti2016mamadroid}.
In fact, we build on it, in that we again model the sequences of (abstracted) calls as Markov chains, and 
use the transition probabilities between states as features.

In order to build the behavioral model in dynamic analysis, we modify a virtual device to allow us to capture the sequence of API calls from the runtime execution trace of apps. 
We call the resulting system \papa, and summarize its operation in \figurename~\ref{fig:diagram}.
First, we execute apps in a virtual device, stimulated by either an automated program (Monkey) or a human. 
We then parse the traces generated by the executions, and extract features for classification. 
The rest of this section presents the details of each component.

\subsection{Virtual Device}\label{sec:virtual}
As mentioned above, the first step in \papa\ is to execute apk samples in a virtual device with either (i)~human users or (ii)~an UI automation tool like Monkey~\cite{monkey}. 
Our virtual device testbed, described in detail below, builds on \chimp, an Android testing system recently presented in~\cite{mario2018chimp} which can be used to collect human inputs from mobile apps.

\descr{\chimp~\cite{mario2018chimp}.} \chimp virtualizes Android devices using the Android-x86 platform, running behind a QEMU instance on a Linux server.
Although it uses an x86 Android image, \chimp actually supports two application binary interfaces (ABI), i.e., both ARM 
and x86 instruction sets are supported.
Once running, the virtualized device can be stimulated by either a locally running automated tool (e.g., Monkey), or the UI can be streamed to a remote browser, allowing humans to interact with it.
\chimp can be used to collect a wide range of data (user interactions, network traffic, performance, etc.) as well as explicit user feedback; however, for the sake of \papa, we modify it to generate and collect \emph{run-time traces}, i.e., %
the call graph of an app's interactive execution.

\descr{Modifications to \chimp.} To effectively monitor malware execution, we substantially modify \chimp from the prototype presented in~\cite{mario2018chimp}, 
which was primarily designed to enable large-scale, human testing of benign apps.
In fact, the original prototype supports code instrumentation via a Java code coverage library called EMMA\longVer{\footnote{\url{http://emma.sourceforge.net/}}}, %
unfortunately, EMMA requires an app's source code to be instrumented, which is %
often not accessible for closed-source apps such as those analyzed in our work.
Therefore, we modify \chimp to get access to debug level run-time information from un-instrumented code.
Note that in Android, each app runs on a dedicated VM which opens a debugger port using Java's Debug Wire Protocol (JWDP).
As long as the \emph{device} is set as debuggable (\texttt{ro.debuggable} property), we can connect to the VM's JWDP port to activate VM level \emph{method tracing}.

We also have to activate tracing: in Android, one can either use Android's Activity Manager (AM) or the DDM Service.
Both end up enabling the same functionality -- i.e., \texttt{startMethodTracing} on \texttt{dalvik.system.VMDebug} and \texttt{android.os.Debug} -- but through different approaches.
That is, AM (via \texttt{adb} \texttt{am}) exposes a limited API that eventually reaches the app via Inter-Process Communication (IPC), while the DDM Service (DDMS, as used by Android Studio) opens a connection directly to the VM's debugger, providing fine grain control over the tracing parameters.
We choose the second approach since it is parameterizable, allowing us to set the trace buffer size, which by default (8MB) can only hold a few seconds of method traces.
Hence, we implement a new DDM Service in \chimp, using the ddmlib library~\cite{ddmlib} to communicate with the VMs and activate tracing. %
Our DDM service multiplexes all tracing requests through a single debugger and we further modify the ddmlib tracing methods to dump traces to the VM file system, and set the trace buffer size to 128MB.
However, apps tested on the virtual device can generate more than 128MB of traces, thus, we add a background job that retrieves and removes traces from the VMs every 30s.
Besides preventing the tracing buffer from filling up, this lets us capture partial traces for apps that might crash during stimulation.

\longVer{To deal with malware masquerading as legitimate applications by using the same package names, we uniquely identify apps based on a hash of the binary, salted with the package name and a unique testing campaign identifier.
This is then mapped to the app's storage location on disk.
We also modify \chimp to disable app verification.}

\subsection{App Stimulation} 
As mentioned, to stimulate the %
AUA, we use both Monkey
and humans.

\descr{Monkey~\cite{monkey}.} Monkey is Android's de-facto standard UI automation tool used to generate inputs.
In \papa, ``Monkeys'' (i.e., more than one Monkey instance) are deployed on the same machine that the virtual devices are running on. %
We set Monkeys to run a single app for 5 minutes (one virtual device VM per app): each Monkey is setup to generate events every 100ms and ignore timeouts, crashes, and security exceptions (although we still log and process them). Setting Monkey to generate input for 5 minutes only should not adversely affect code coverage, as prior work~\cite{choudhary2015automated} reports that most input generators achieve maximum coverage between 5 to 10 minutes.
As Monkey may generate events at a higher frequency than some apps can process, we also re-run offending apps with a decreased rate (300ms).
As discussed in Section~\ref{sec:preprocessing}, some apps fail to execute, for one of three reasons: (i)~they fail to install, (ii)~crash, or (iii)~have no interactive elements (e.g., background apps), as observed through \texttt{logcat} and from the Monkey output itself.

\descr{Humans.} In order to have real users stimulate the samples, we recruited about 5k workers (5,030) from the Crowdflower.com crowdsourcing platform that are ``historically trustworthy''.
We let them interact with the virtual device by streaming its UI to their browser via an HTML5 client.
The client transmits user actions back to \papa, which translates them to Android inputs and forwards them to the virtual device.
In addition to the virtual device UI, user controls were provided to, e.g., move to the next app in the testing session.

\begin{figure}[t]
 \center
 \subfigure[\label{fig:users}]
 {\includegraphics[width=0.238\textwidth]{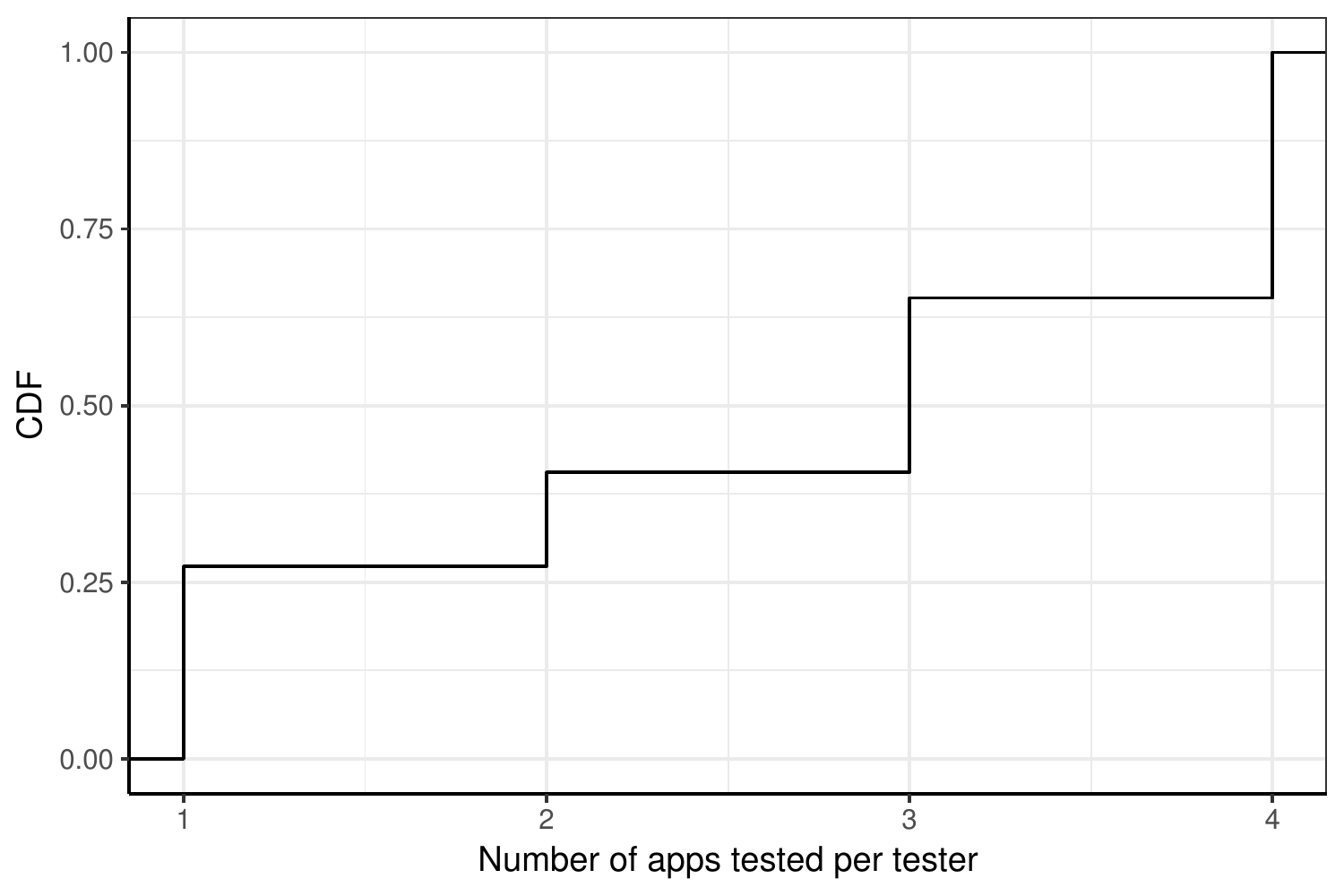}}
 \subfigure[\label{fig:appruns}]
{\includegraphics[width=0.238\textwidth]{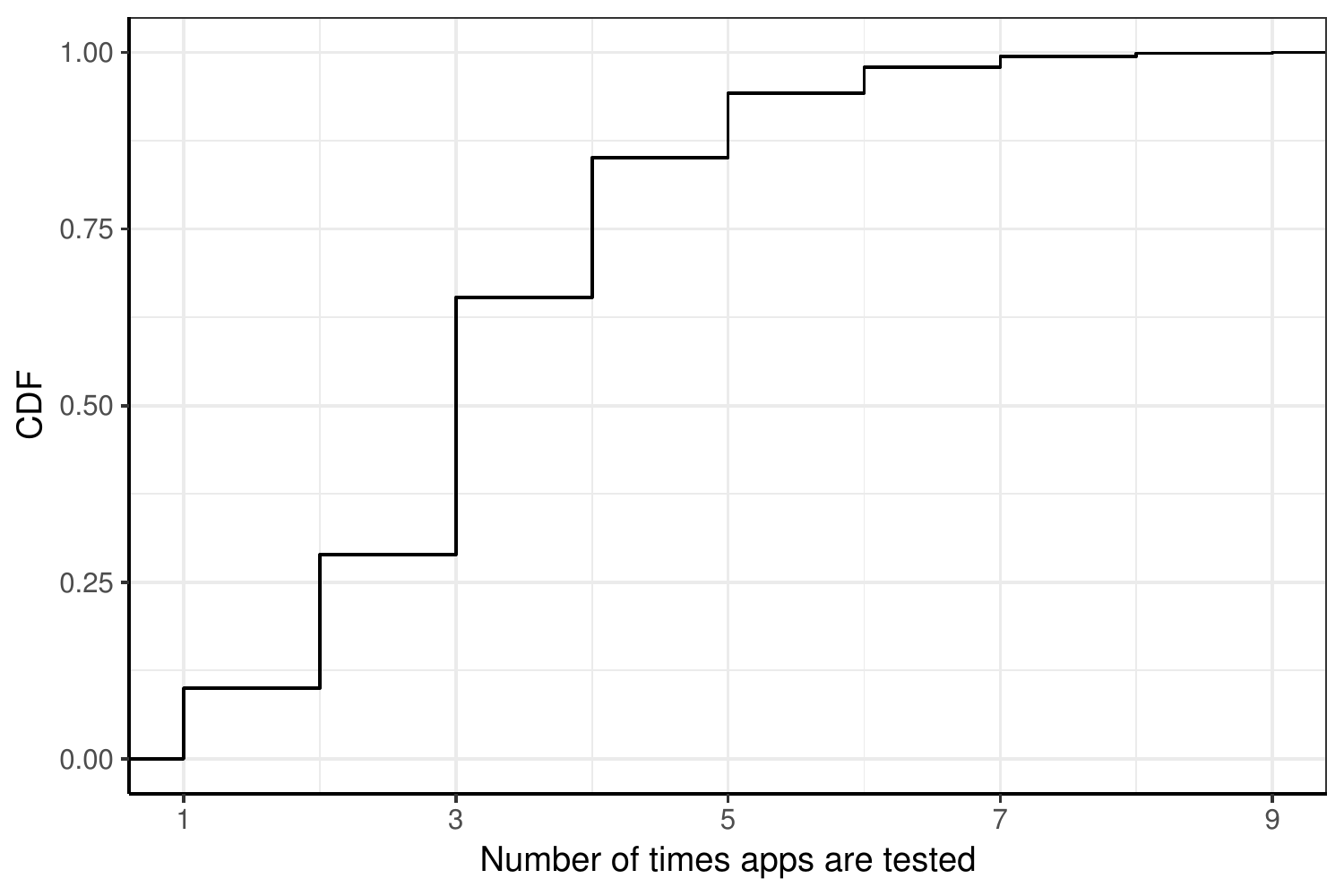}}
 \caption{Cumulative distribution function of the (a) number of apps tested per tester, and (b) number times apps are tested.}
\end{figure}

Each user is given 4 randomly selected apps from our dataset and told to explore as much of each app's functionality as possible before proceeding to the next app.
CHIMP already provides heuristics to discard users with low engagement, and we do not enforce a lower bound on the time users must spend testing apps, since given the nature of our sample, some apps might have limited interaction opportunities.
Consequently, we aim to have a median of at least three different users stimulate each app. 
In \figurename~\ref{fig:users} and~\ref{fig:appruns}, respectively, we plot the CDF of the number of apps each user tests and the number of times an app is tested. 
We also limit app install time to 40s to avoid frustrating the users.
We run the test sessions between August 9th and 11th, 2017 and we pay each user \$0.12 per session.

\descr{Ethics.} For the experiments involving humans, we have obtained approval through our institution's ethical review process.
Although we requested basic demographic data (age, gender, country), %
we did not collect privacy sensitive information, and users were instructed not to enter any real, personal information, e.g., account details. 
\longVer{No software had to be installed on participants' devices: they interacted with the apps on a webpage, while we collected information about the app execution in the background.}
Also note that we provided email credentials to use when required, so that they did not have to use their own credentials or other contact information.

\subsection{Trace Parsing}
\label{tracing}

As discussed above, our virtual device component takes care of collecting method traces, network packets, and event logs generated when the app is running. 
To parse these traces, one could use different strategies,
for instance, tracking data flow from selected sources (e.g., the device id -- {\tt getDeviceID()}) to sinks (e.g., a data output stream -- {\tt writeBytes()}), or using frequency analysis to derive commonly used API calls by malware (as in DroidAPIMiner~\cite{Aafer2013DroidAPIMiner}).

\papa follows the behavioral model based approach of \mama, based on the {\em sequences} of API calls that the app performs at runtime, %
rather than statically extracting it from the apk.
This way, we aim to capture different behavior when benign and malicious apps invoke API calls. %
For instance, a benign SMS app might receive an SMS, get the message body using {\tt getMessageBody()} and afterwards, display the message to a user via a view by executing, in sequence, {\tt setText(String msg)} and {\tt show()} methods of the view.
A malicious app, however, might exfiltrate all received SMSs by executing {\tt sendTextMessage()} for every message before displaying it.

\begin{figure}[t]
 \begin{center}
\resizebox{0.4\textwidth}{!}{\scriptsize
\begin{tikzpicture}[every node/.style={draw},text height=0.1cm,text width=3.80cm,align=center]
\draw(0,0) node (A0) {air.com.eni.ChefJudy030\\.AppEntry.onNewIntent};
\draw(0.5,-0.5) node [draw=none] {\hspace{-1cm} Trace 1};
\draw[very thick] (0.3,-1) -- (0.6, -1);
\draw[very thick] (0.45,-0.85) -- (0.45, -1.15);
\draw(0,-1.75) node (A1) {air.com.eni.ChefJudy030\\.AppEntry.onNewIntent};
\draw(0.5,-2.25) node [draw=none] {\hspace{-1cm} Trace 2};
\draw[very thick] (0.3,-2.75) -- (0.65, -2.75);
\draw[very thick] (0.3,-2.85) -- (0.65, -2.85);
\draw(0,-4) node (A6) {air.com.eni.ChefJudy030\\.AppEntry.onNewIntent};
\draw(0.5,-4.5) node [draw=none] {\hspace{-1cm} Aggregated Trace};
\draw(4.75,1.25) node (A2) {air.com.eni.ChefJudy030\\.AppEntry.InvokeMethod};
\draw(4.75,0.5) node (A3) {java.lang.Class.getMethod};
\draw(4.75,-0.25) node (A4) {android.app.Activity\\.onNewIntent};
\draw(4.75,-1) node (A5) {air.com.eni.ChefJudy030\\.AppEntry.InvokeMethod};
\draw(4.75,-1.75) node (A7) {java.lang.Class.getMethod};
\draw(4.75,-2.5) node (A8) {android.app.Activity\\.onNewIntent};
\draw(4.75,-3.25) node (A9) {air.com.eni.ChefJudy030\\.AppEntry.InvokeMethod};
\draw(4.75,-4) node (A10) {java.lang.Class.getMethod};
\draw(4.75,-4.75) node (A11) {android.app.Activity\\.onNewIntent};
\draw[->,thick] (A0) -- (A2.west) node [draw=none, midway=10pt] {\hspace{-1cm}3};
\draw[->,thick] (A0) -- (A3.west) node [draw=none, midway=10pt] {\hspace{-1cm}3};
\draw[->,thick] (A0) -- (A4.west) node [draw=none, midway=10pt] {\hspace{-1cm}3};
\draw[->,thick] (A1) -- (A5.west) node [draw=none, midway=10pt] {\hspace{-1cm}1};
\draw[->,thick] (A1) -- (A7.west) node [draw=none, midway=10pt] {\hspace{-1cm}1};
\draw[->,thick] (A1) -- (A8.west) node [draw=none, midway=10pt] {\hspace{-1cm}1};
\draw[->,thick] (A6) -- (A9.west) node [draw=none, midway=10pt] {\hspace{-1cm}4};
\draw[->,thick] (A6) -- (A10.west) node [draw=none, midway=10pt] {\hspace{-1cm}4};
\draw[->,thick] (A6) -- (A11.west) node [draw=none, midway=10pt] {\hspace{-1cm}4};
\end{tikzpicture}
}
 \caption{Aggregated sequence of API calls showing the direct children of call {\tt air.com.eni.ChefJudy030.AppEntry.onNewIntent}, and the number of times they are called (numbers on the arrow).}
 \label{fig:sequence}
 \end{center}
\end{figure}
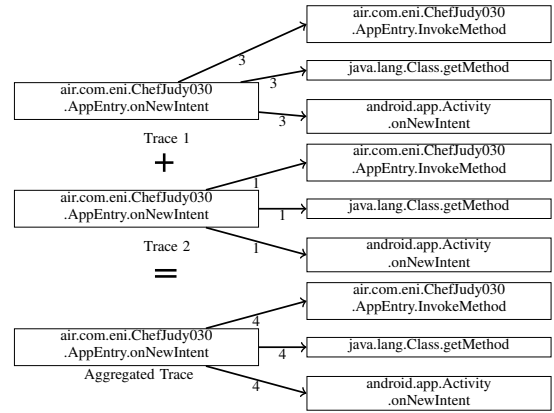

To derive the API call sequences, we collect the method traces and transform them into a call graph using {\em dmtracedump}~\cite{dmtracedump}.
From the call graph, we then extract the sequences using a custom script, while preserving the %
number of times an API call is %
executed as a multiplier in each sequence. %
As discussed above, to avoid losing traces when the trace buffer is full, we collect virtual device traces every 30s, and clear the buffer for incoming traces. Along the same lines, we have a median of three different users run the same app to improve the quality of the traces gathered. As a result, we aggregate the sequences of API calls they generate for the same app into a single sequence. 
In \figurename~\ref{fig:sequence}, we provide an example of the sequence for the API call {\tt air.com.eni.ChefJudy030.AppEntry.onNewIntent} when aggregated from two other sequences. We do not show the params and return type to ease presentation. Also, in some cases, {\em Trace 1} may contain calls in a sequence that is not called in {\em Trace 2}, hence, the aggregated trace also reflects such calls. %

\subsection{Feature Extraction}

As in \mama~\cite{mariconti2016mamadroid}, which operates in one of two modes i.e., {\tt family} or {\tt package}, \papa also abstracts each API call in the parsed trace to its corresponding family and package names using the Android API packages from API level 26 and the latest Google API packages.
The abstraction allows for resilience to API changes in the Android framework as packages are added or deprecated less frequently compared to single API calls. It also helps to reduce the feature set size as the feature vector of each app is the square of the number of states in the Markov chain. %

Also note that we  modify \mama's method of abstracting API calls: before performing abstraction to packages or families, we first abstract an API call to its class using a whitelist approach. We do this to avoid abstracting an API call to the wrong package or family in case an app prefixes its package name with one from the Android or Google APIs.

We then build a \emph{behavioral model} from the %
abstracted sequences of API calls by building a Markov chain that models the sequences from which we extract features  %
used to classify an app as either benign or malicious. 
More specifically, features are %
the probability of transitioning from one state (i.e., API call) to another in the Markov chain representing the API call sequences in an app. 
%

%

%
%
%
%
%
%

%
%

\subsection{Classification}
Finally, we perform classification, labeling an app as benign or malware using a supervised machine learning classifier. 
More specifically, we use Random Forests with 10-fold cross validation. 
We choose Random Forests since it performs well on binary classification over high-dimension feature spaces, which is the case with the Markov chain based model from \mama.

\section{Experimental Setup}\label{sec:setup}
In this section, we introduce our experiments, the datasets used in our evaluation, as well as the preprocessing carried out on them.

\subsection{Overview of the experiments}
As discussed in Section~\ref{sec:intro}, we aim to perform a comparative analysis of Android malware detection systems based on behavioral models, using static, dynamic, and hybrid analysis.
To this end, we perform three sets of experiments.
(1) {\em Static:} We  evaluate \mama, which performs Android malware detection based on behavioral modeling in static analysis; %
(2) {\em Dynamic:} We analyze the detection performance of \papa (see Section~\ref{sec:papaeval}), which uses dynamic analysis, while also comparing automated input generation (Monkey) and human-generated input; %
(3) {\em Hybrid:} We combine static and dynamic analysis by merging the sequences of API calls from both methods, once again comparing Monkey and human based input generation.

All methods operate in one of two levels of abstraction, i.e., API calls are abstracted to either their family or package names. 
Overall, we use the same modeling technique and the same machine learning classifier.
More specifically, we use the Random Forests classifier and in {\tt family} (resp., {\tt package}) mode, we use a configuration of 51 (resp., 101) trees with depth equal to 8 (resp., 32).

\subsection{Datasets}\label{sec:dataset}
Our evaluation uses two datasets: \longVer{a collection of} recent malware samples and a dataset of random benign apps, as discussed below.

\descr{Benign Samples.} For consistency, we opt to re-use the set of 2,568 benign apps labeled as ``newbenign'' in\longVer{ the \mama paper}
~\cite{mariconti2016mamadroid}.
In June 2017, we re-downloaded all the apps in order to ensure we have working apps and their latest version,
obtaining 2,242 (87\%) apps. %
We complement this list with a 33\% sample of the top 49 apps (as of June 2017) from the 29 categories listed on the Google Play Store,
adding an additional 481 samples.
Overall, our benign dataset includes a total of 2,723 apps.

\descr{Malware Samples.} Our malware dataset includes samples obtained in June 2017 from VirusShare -- a repository of apps that are likely to be
malicious.\footnote{\url{https://virusshare.com/}} 
More precisely, VirusShare contains samples that have been detected as malware on various OS platforms, including Android.
To obtain only Android malware, we check that each sample is correctly zipped and packaged as an apk, contains a Manifest file, and has a package name.
Using this method, we gather 2,692 valid Android samples labeled as malware in 2017 by the antivirus engines on VirusShare.
In addition, we add two more apps (Chef Judy and Dress Up Musa) from the Google Play Store reported as malware in the news and later removed from the play store.\footnote{\url{https://goo.gl/hBjm0T} and \url{https://goo.gl/IQprtP}} 
In total, our malware dataset includes 2,694 apps.

\subsection{Data Pre-Processing}\label{sec:preprocessing}

\descr{Static Analysis.} For static analysis, we re-use the source code of \mama available on bitbucket.
We set a timeout limit of six hours for call graph extraction, and are unable to obtain the call graphs for 98 (3.6\%) and 251 (9.3\%) apps in the benign and malware datasets, respectively. %
This is consistent with experiments reported in~\cite{mariconti2016mamadroid}, due to the timeout but also to samples exceeding memory requirement (we allocate %
16GB for the JVM heap space).

\descr{Dynamic Analysis.} During dynamic analysis, before running the apps on the virtual device,
we process them statically using androguard\footnote{\url{https://github.com/androguard/androguard}} to determine whether they have activities. Out of the total 5,417 apps in our datasets, we find that 82 apps contain no activity.
As interaction with an Android app requires visuals that users can click, tap, or touch to trigger events, we therefore exclude these from the samples to be stimulated using Monkey or humans. 
We also remove 244 apps that do not have a launcher activity, since launching these apps on the virtual device will have no visual effect; i.e., no UI will be displayed to the tester.
Finally, we do not include 139 apps which fail to install on the virtual device for one of the reasons shown in Table~\ref{table:install}.

\descr{Hybrid Analysis.} 
To obtain a hybrid detection system, we merge the sequences of abstracted API calls (from which we extract features) obtained using both static and dynamic analysis. More specifically, we merge the sequences of API calls following the same strategy used to aggregate the traces discussed in Section~\ref{tracing}. 
Naturally, for hybrid analysis, we use samples %
for which we have traces for both static and dynamic analysis.

\begin{table}[t]
\centering
\setlength{\tabcolsep}{0.2em} %
\resizebox{0.95\linewidth}{!}{
\begin{tabular}{lrr}
\toprule
{\bf Failure}  & {\bf Benign} & {\bf Malware} \\
\midrule
Already installed & 10 & 9 \\ %
Contains native code not compatible with the device's CPU & 0 & 2\\
App's dex files could not be optimized and validated & 0 & 1\\
Apk could not be unarchived by Android aapt & 3 & 4\\
Shared library requested by app is not available on the device & 0 & 1\\
Does not support the SDK (version 4.4.2) on the device & 36 & 6 \\
Requests a shared user already installed on the device & 0 & 1\\
Android's failure to parse the app's certificate & 0 & 4\\
Fails to complete installation within time limit (40s) & 39 & 23\\
\midrule
{\bf Total: } & {\bf 88} & {\bf 51}\\ %
\bottomrule
\end{tabular}
}
\vspace{-0.1cm}
\caption{Reasons why apps fail to install on the virtual device.} 
\label{table:install}
\vspace{-0.1cm}
\end{table}

\descr{Final Datasets.} In Table~\ref{table:evaldata} we report, in the right-most column, the final number of samples in each dataset, %
for each method of analysis. 
During dynamic analysis, we fail to obtain traces for 724 apps when stimulating with Monkey and 693 when stimulating with humans. 
We discuss the reasons for the failure after examining the logs, in Appendix~\ref{challenges}. 
This happens for various reasons, and we defer further analysis to the full version of the paper
Note that the hybrid analysis method consists of samples for which we obtain traces both statically and dynamically.

\begin{table}[t]
\centering
\footnotesize
\setlength{\tabcolsep}{3pt}
\resizebox{0.815\columnwidth}{!}{
\begin{tabular}{llrrr}
\multicolumn{5}{c}{}\\
\toprule
{\bf Analysis} & {\bf Stimulator} & {\bf Category} & {\bf \#Samples} & {\bf \#Traces /}\\
& & & & {\bf Call graphs} \\
\midrule
{\bf Static} & -- & Benign & 2,723 & 2,625 \\
(\mama) & & Malware & 2,694 & 2,443 \\
  \midrule
{\bf Dynamic} & {Human} & Benign & 2,596 & 2,348 \\
(\papa) & & Malware & 2,356 & 1,911 \\
 & {Monkey} & Benign & 2,596 & 2,336 \\
 & & Malware & 2,356 & 1,892 \\
\midrule
{\bf Hybrid} & {Static \& Human} & Benign & 2,596 & 2,235 \\
 & & Malware & 2,356 & 1,708 \\
	& {Static \& Monkey} & Benign & 2,596 & 2,234 \\
 & & Malware & 2,356 & 1,686 \\
  \bottomrule
\end{tabular}
}
\vspace{-0.1cm}
\caption{Datasets used to evaluate each method of analysis. %
} 
\vspace{-0.2cm}
\label{table:evaldata}
\end{table}

\section{Evaluation}
\label{sec:eval}
We now present the results of our experiments, reporting detection performance and, for dynamic analysis, code coverage.

\subsection{Static Analysis}\label{sec:static}
To evaluate the static analysis technique, we use a slightly modified version of \mama~\cite{mariconti2016mamadroid}. %
Also note that, while~\cite{mariconti2016mamadroid} uses API level 24, we use the more recent API level 26.
We run our experiments on the %
samples (2,625 benign and 2,443 malware) for which we obtain call graphs, 
and report the $F$-measure obtained when operating in {\tt family} 
and {\tt package} modes in the top two rows of Table~\ref{table:allresults}.
We observe that the latter performs slightly better, achieving $F$-measure of 0.91, compared to 0.86 in the former %
which is consistent with the results reported in~\cite{mariconti2016mamadroid}. %
The {\tt package} mode achieves higher $F$-measure than {\tt family} mode as it captures the behavior of apps at a finer granularity which reveals more distinguishing behaviors between malware and benign apps as demonstrated by higher precision and recall (see Table~\ref{table:allresults}).

\begin{table}[t]
\centering
\footnotesize
\setlength{\tabcolsep}{2.5pt}
\resizebox{0.85\columnwidth}{!}{
\begin{tabular}{lllrrr}
\toprule
{\bf Analysis} & {\bf Stimulator} & {\bf Mode} & {\bf ${F}$-measure} & {\bf Precision} & {\bf Recall}\\
\midrule
{\bf Static} & -- & {\tt Family} & {\bf 0.86} & 0.84 & 0.88 \\
 (\mama) & & {\tt Package} &  {\bf 0.91} & 0.89 & 0.93 \\
 \midrule
 {\bf Dynamic} & Human & {\tt Family} & {\bf 0.85} & 0.80 & 0.90 \\
 (\papa) & & {\tt Package} &  {\bf 0.88} & 0.84 & 0.92  \\[0.75ex]
   & Monkey & {\tt Family} & {\bf 0.86} & 0.84 & 0.89 \\
 & & {\tt Package} &  {\bf 0.92} & 0.91 & 0.93 \\
 \midrule
 {\bf Hybrid} & Static \& Human & {\tt Family} & {\bf 0.87} & 0.86 & 0.88  \\
 & & {\tt Package} &  {\bf 0.90} & 0.88 & 0.91 \\ [0.75ex]
 & Static \& Monkey & {\tt Family} & {\bf 0.88} & 0.88 & 0.89  \\
 & & {\tt Package} &  {\bf 0.92} & 0.92 & 0.93 \\
   
  \bottomrule
\end{tabular}
}
\vspace{-0.1cm}
\caption{Results achieved by all analysis methods while %
using human and Monkey as app stimulators during dynamic analysis.} %
\label{table:allresults}
\vspace{-0.3cm}
\end{table}

\subsection{Dynamic Analysis}\label{sec:papaeval} 
Next, we report the results achieved by dynamic analysis (i.e., using \papa), comparing between stimulation performed by Monkey and humans.

\descr{Detection Performance.} %
For Monkey, we use the dataset shown in Table~\ref{table:evaldata}, i.e., on %
2,336/1,892 samples for benign/malware. When \papa runs in {\tt family} mode, it achieves $F$-measure, precision, and recall of 0.86, 0.84, and 0.89, respectively. Whereas in {\tt package} mode, it achieves $F$-measure, precision, and recall of 0.92, 0.91, and 0.93, respectively, as reported in Table~\ref{table:allresults}. 
When humans stimulate the apps (2,348 benign and 1,911 malware) and \papa runs in {\tt family} mode, we get $F$-measure, precision, and recall of 0.85, 0.80, and 0.90, respectively. Whereas when operating in {\tt package} mode, $F$-measure, precision, and recall go up to 0.88, 0.84, and 0.92, respectively (see Table~\ref{table:allresults}). 

Overall, lower $F$-measures in all modes of operation in dynamic analysis compared to static analysis (i.e., \papa vs \mama) are due to  increases in false positives.
In fact, recall is  around 0.90 on all experiments, while precision is as low as 0.80 ({\tt family} mode with humans).

\longVer{
\begin{figure}[t]
 \center
 \subfigure[\label{fig:coverage}]
 {\includegraphics[width=0.238\textwidth]{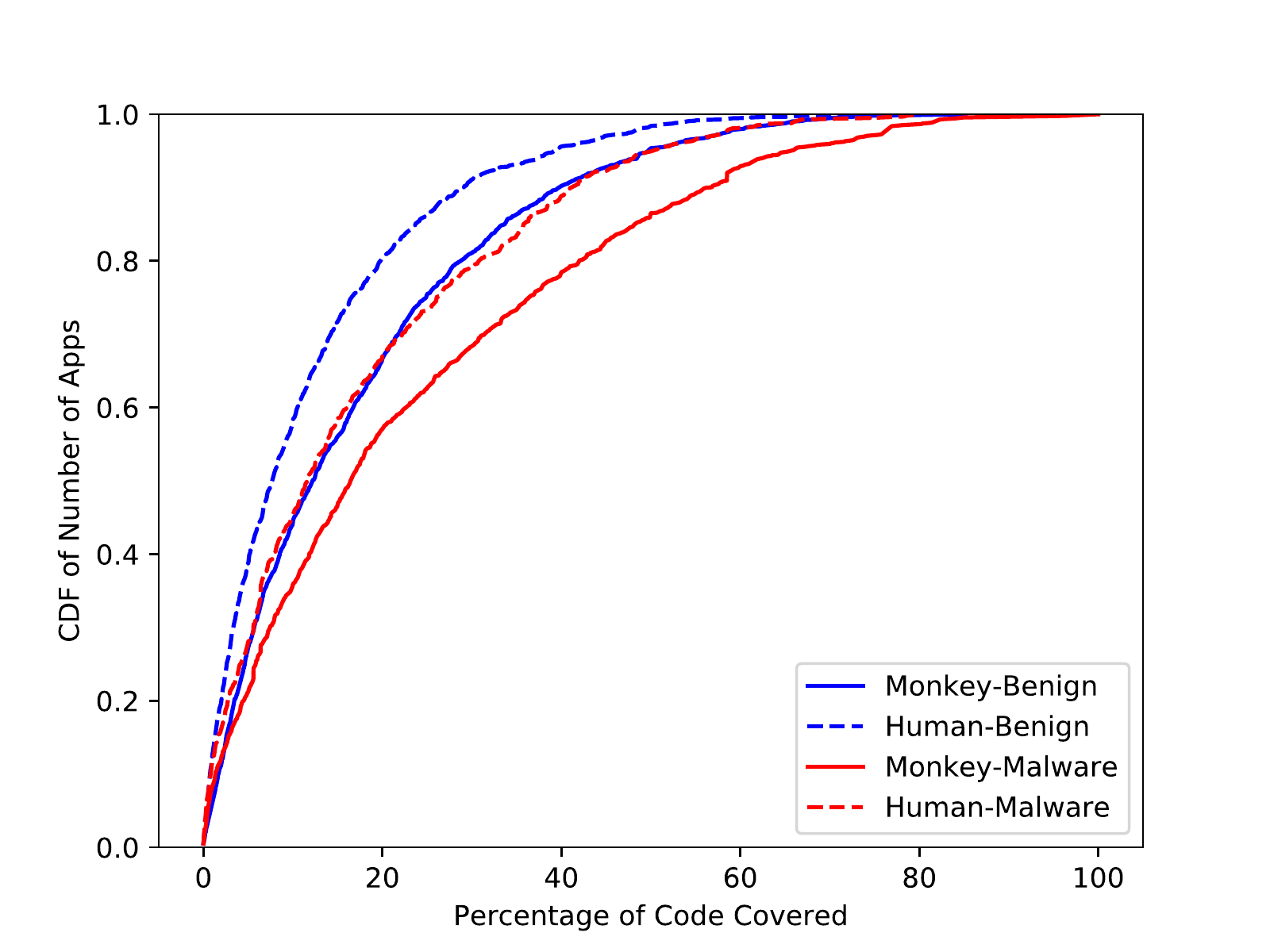}}
  \hspace{-0.15cm}
 \subfigure[\label{fig:dynamicCode}]
{\includegraphics[width=0.238\textwidth]{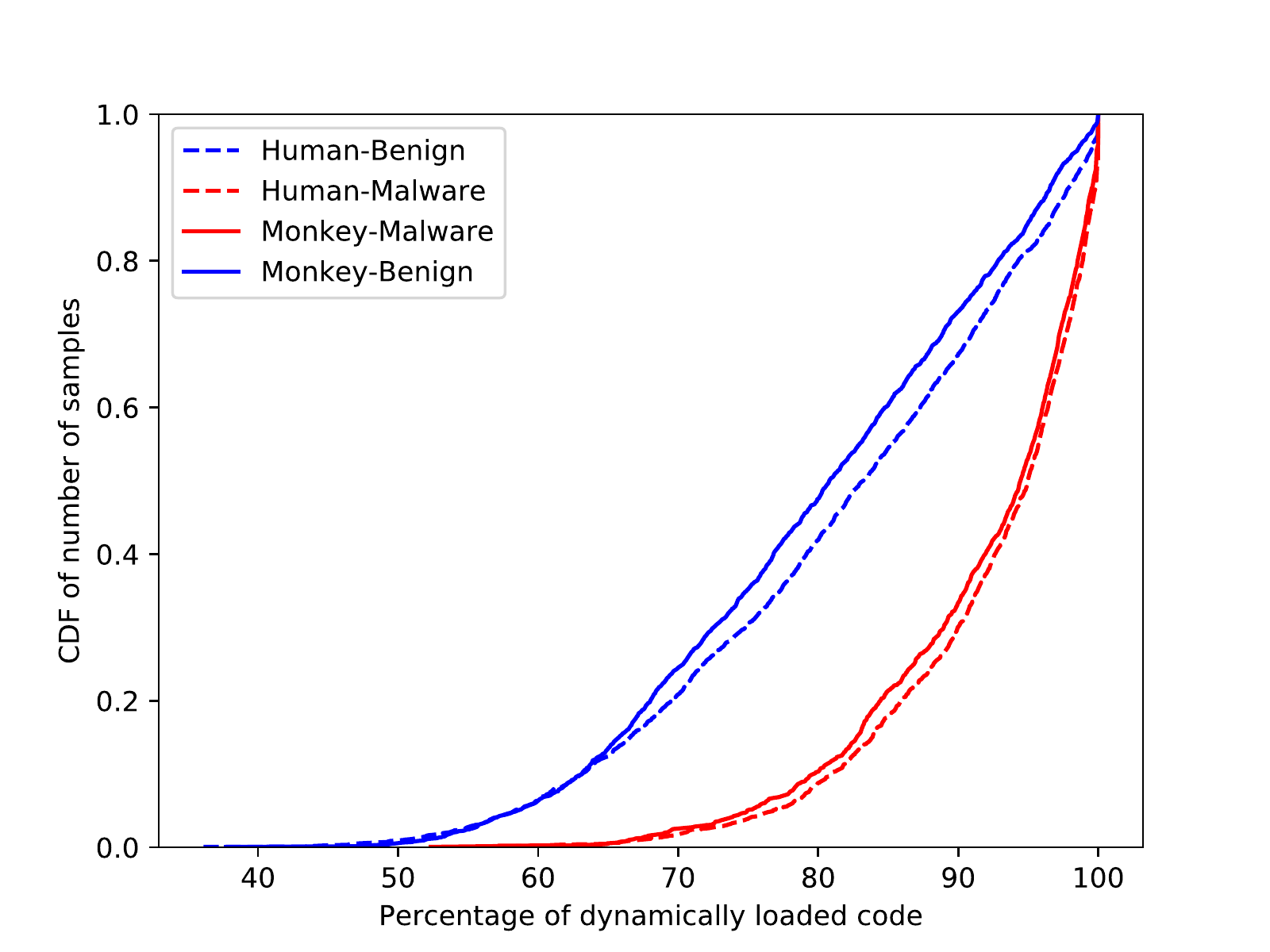}}
\vspace{-0.4cm}
 \caption{Cumulative distribution function of the percentage of (a) code covered in benign and malicious apps when they are stimulated by Monkey and human, and (b) API calls that are dynamically loaded during dynamic analysis.}
\vspace{-0.2cm}
\end{figure}}

\shortVer{
\begin{figure}[t]
 \center
 \includegraphics[width=0.24\textwidth]{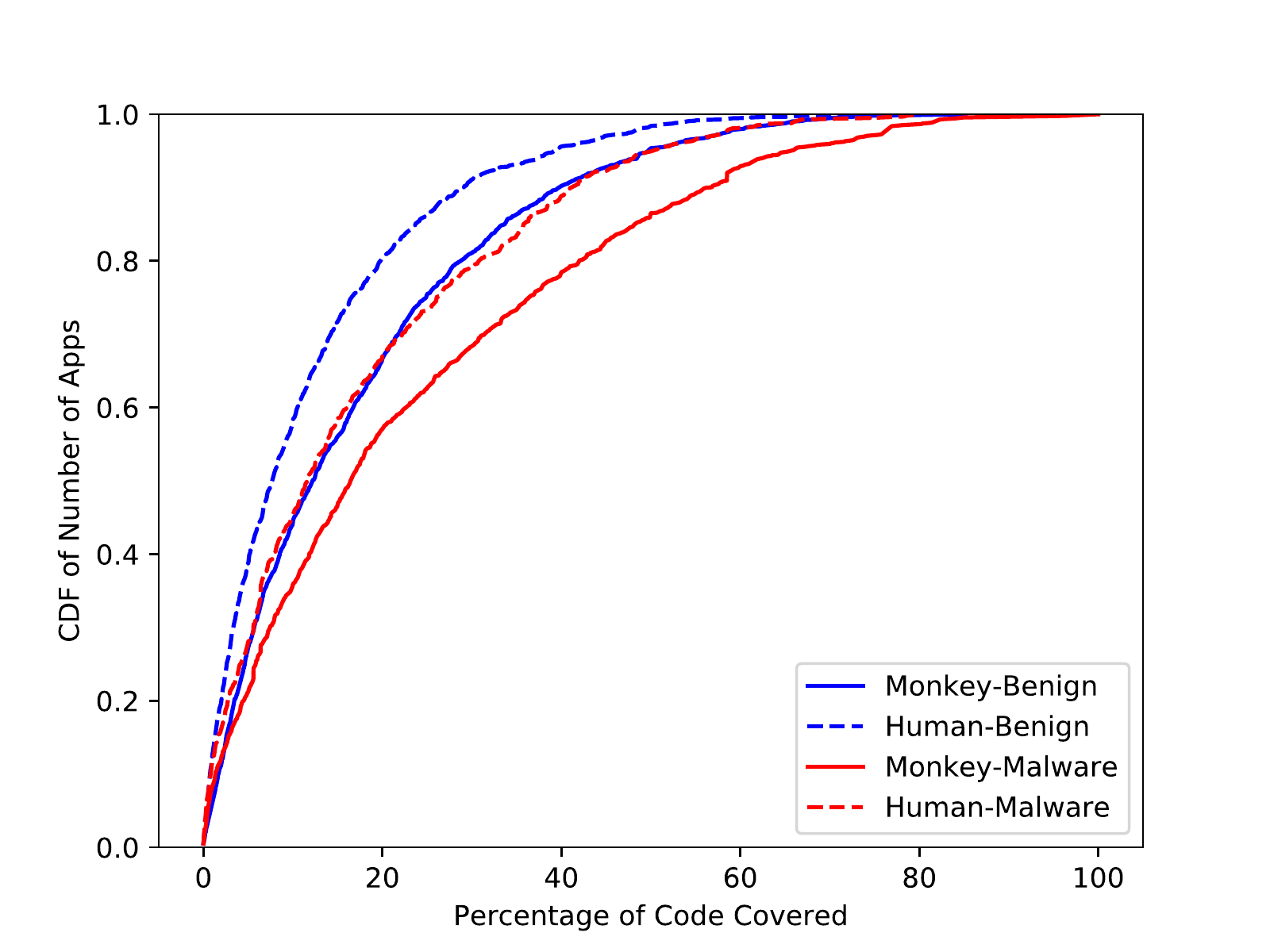}
 \caption{Cumulative distribution function of the percentage of code covered in benign and malicious apps when they are stimulated by Monkey and human.}
\vspace{-0.2cm}
\label{fig:coverage}
\end{figure}
}

\descr{Code Coverage.} As mentioned, the performance of dynamic analysis tools is affected by whether malicious code is triggered during execution. Since \papa relies on the sequences of API calls to detect malware, we analyze  code coverage of each app to measure how much of an app's API calls Monkey/humans successfully trigger. Thus, we focus on API calls that begin with the package name of the app.

For the apps for which we obtain traces %
(85\% and 86\%, resp., for Monkey and human), Monkey is able to trigger on average, 20\% of the API calls. %
In \figurename~\ref{fig:coverage}, we plot the cumulative distribution function (CDF) of the percentage of code covered, showing that for 90\% of the benign apps, at least 40\% of the API calls are triggered by Monkey. Whereas with respect to the malware samples, at least 57\% 
of the API calls are triggered. %
As for humans, we find that users are able to trigger, on average, 14\% of the API calls. %
Similarly, \figurename~\ref{fig:coverage} shows that at least 29\% %
of the API calls %
are triggered in 90\% of the benign apps. %
However, with 90\% of the malicious apps, 41\% %
of the API calls %
are triggered. %

With both stimulators, there is a higher percentage of code coverage in the malware apps than in benign apps. This is due to malware apps being smaller in size compared to the benign apps in our dataset. The mean number of API calls in the benign and malware apps are respectively, 43,518 and 16,780. However, with respect to stimulators, Monkey is able to trigger more code in apps compared to humans,
which is likely due to Monkey triggering more events than humans in the time each spend testing the apps.

\shortVer{We also investigate the prevalence of dynamic code loading in the wild, as it could be used for malicious purposes~\cite{poeplau2014execute} (e.g., to evade static analysis). 
Due to space limitations, we defer findings to the full version of the paper~\cite{arxiv}.}

\longVer{\descr{Dynamic code loading.} We also report the percentage of code that is dynamically loaded by apps by examining the API calls that exist in the dynamic traces but not in the apk. That is, we extract the API calls from the apk using apktool~\cite{apktool} and measure the percentage of the calls that are in the dynamic traces but not in that extracted statically. We do this to evaluate the prevalence of dynamic code loading in apps as this could be used for malicious purposes~\cite{poeplau2014execute}. For instance, malware developers may design their apps (or repackage benign apps) to only dynamically load the malicious code and evade static analysis tools~\cite{Zhou2012dissecting,suarez2018eight}.  
In \figurename~\ref{fig:dynamicCode}, we report the CDF of the percentage of the code that is dynamically loaded, finding that in 90\% of the samples, 97\% of the API calls are dynamically loaded irrespective of the app stimulator. 
As for malware and benign apps, about 99.5\% of the API calls are dynamically loaded in 90\% of malware samples irrespective of the stimulator compared to about 97\% in benign apps. We also evaluate the percentage of the dynamically loaded code that is common to all apps to measure whether they are primarily virtual device or OS operations such as app installation, start/stop method tracing, os handler etc. We find that only 0.14\% and 0.08\%, respectively, are common across apps when the stimulator is human and Monkey. %

Overall, with up to 99.5\% of code being dynamically loaded in 90\% of our malware samples, we believe that dynamic code loading might indeed pose a problem for malware detection tools based solely on static analysis and dependent on API calls. That is, these tools might not be resilient to evasion techniques that load code dynamically, since a large portion of the executed code will not be examined at all. On the other hand, with benign apps also heavily employing dynamic code loading, we cannot conclude that only apps that load a large portion of their code during dynamic analysis are malicious.}

\subsection{Hybrid Analysis}
We now report the results achieved by hybrid analysis, comparing between stimulation performed by Monkey and humans. Recall that only samples for which we have obtained a trace in both static and dynamic analysis, as reported in Table~\ref{table:evaldata}, are merged and evaluated.

In {\tt family} mode, the hybrid system, using traces produced by Monkey, achieves an $F$-measure of 0.88, whereas when using traces produced by humans, 0.87. When operating in {\tt package} mode and using Monkey, it achieves  an $F$-measure of 0.92, and 0.90 with humans, as reported in Table~\ref{table:allresults}.

Note that we do not report code coverage in hybrid analysis because the traces from static analysis are an overestimation of the API calls in the app. Hence, merged traces do not reflect code covered in each app when executed.

\section{Comparative Analysis}
\label{sec:comparison}
We now set out to examine and compare: (1) the detection performance of each analysis method, i.e., detecting malware based on a behavioral model built via static, dynamic, or hybrid analysis, (2) the samples that are misclassified in each method, and (3) the samples misclassified in one method but correctly classified by another.
Due to each method having inherent limitations, it is not clear from prior work how they compare against each other. Therefore, in this section, we shed light on their comparisons.

\subsection{Detection Performance}
\label{sec:detect}

\begin{figure}[t]
 \center
 \subfigure[\label{fig:coverageB}]
 {\includegraphics[width=0.238\textwidth]{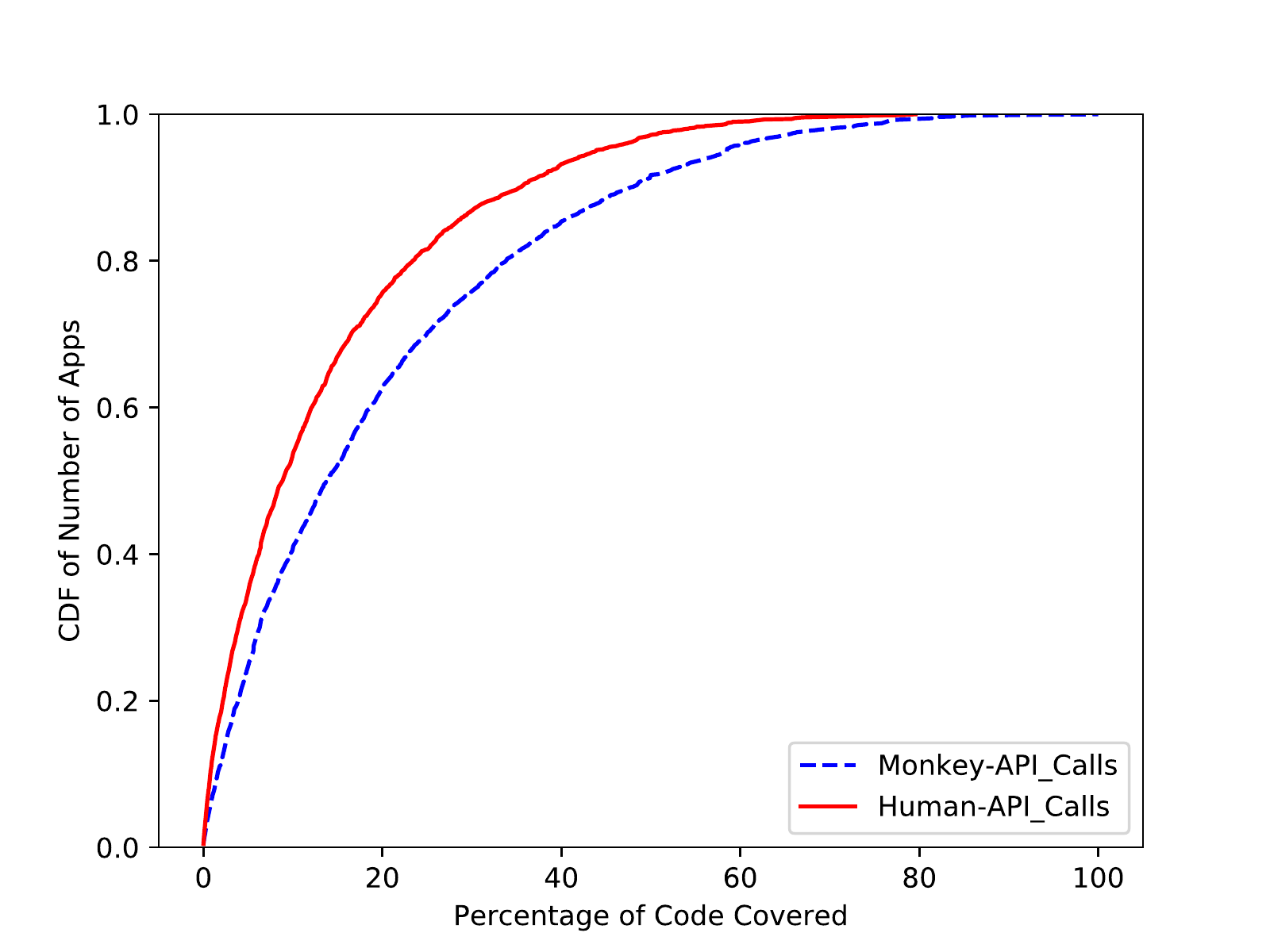}}
 \hspace{-0.15cm}
 \subfigure[\label{fig:coverageM}]
 {\includegraphics[width=0.238\textwidth]{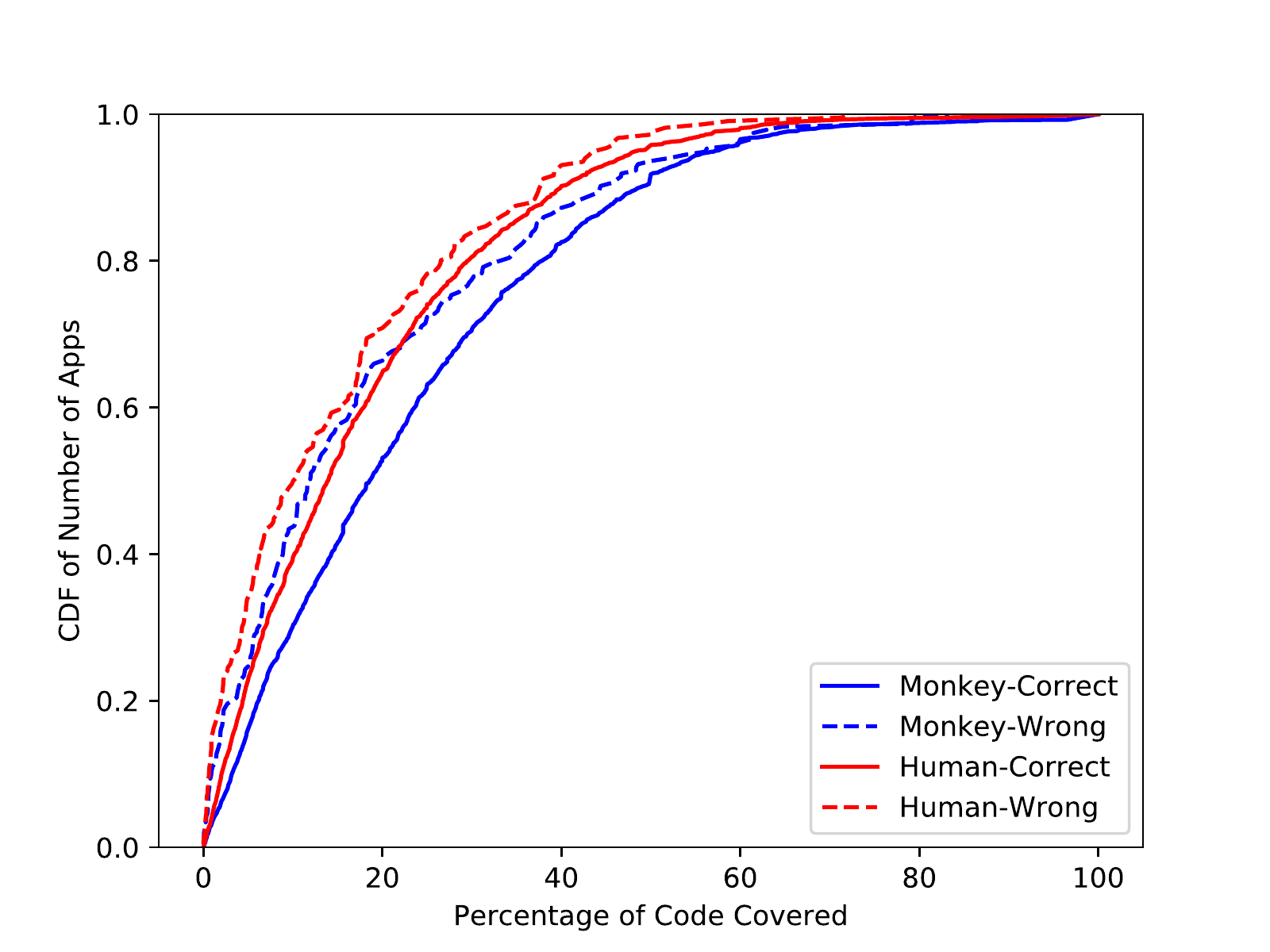}}
 \vspace{-0.25cm}
 \caption{Cumulative distribution function of the percentage of code covered (a) when apps are stimulated by humans and Monkey, and (b) when the correctly classified and misclassified apps are stimulated by humans and Monkey.}
 \vspace{-0.3cm}
\end{figure}

We start by comparing the results of the different analysis methods. 
Recall that we have abstracted 
each API call to either its family or package name, therefore, each %
method operates in one of two modes. When operating in {\tt family} mode, with static analysis we achieve an $F$-measure of 0.86, whereas, with dynamic analysis, we achieve $F$-measure of 0.86 when apps are stimulated by Monkey and 0.85 when stimulated by humans (See Table~\ref{table:allresults}). %
In {\tt package} mode, we achieve $F$-measure of 0.91 with static analysis, whereas with dynamic analysis we achieve $F$-measure of 0.92 when apps are stimulated by Monkey and 0.88 when stimulated by humans (See Table~\ref{table:allresults}). %

The results show that static analysis is at least as effective as dynamic analysis depending on the app stimulator used during dynamic analysis. We believe this is because the behavioral model used to perform detection primarily leverages API calls. Although static analysis is not able to detect maliciousness when code is loaded dynamically, it provides an overestimation of the API call sequences in the apk. Consequently, all behaviors that can be extracted from the apk are actually captured by the static analysis classifier.
On the other hand, dynamic analysis captures only the behavior exhibited by the samples during runtime. Hence, any behavior not observed during runtime is not used in the decision-making of the dynamic analysis classifier. 

To verify this hypothesis, we evaluate how the percentage of code covered differs when different app stimulators are employed as well as in correctly classified and misclassified samples. From \figurename~\ref{fig:coverageB}, we observe that when Monkey is used as the app stimulator, at least 48\% of the API calls are triggered in 90\% of the samples, compared to 35\% when they are stimulated by humans. Similarly, as shown in \figurename~\ref{fig:coverageM}, 49\% of the API calls are triggered in 90\% of the samples correctly classified when Monkey is used to stimulate apps compared to 44\% of API calls in 90\% of the apps that are misclassified. When humans are used to stimulate the apps, 40\% of the API calls are triggered in 90\% of samples that are correctly classified compared to 38\% triggered in 90\% of samples misclassified.
As a result of better code coverage, dynamic analysis performs better when apps are stimulated by Monkey compared to when apps are stimulated by crowdsourced users.
Therefore, we find that, other than the non-susceptibility to evasion techniques such as dynamic code loading, dynamic analysis tools based on API calls may have no advantage over static analysis based tools unless the code coverage is improved.

However, when traces from static and dynamic analysis are merged into a hybrid system, %
in {\tt family} mode, %
we achieve $F$-measure of 0.88 using Monkey compared to 0.86 achieved by both static analysis and dynamic analysis (with Monkey) alone. Similarly, we achieve $F$-measure of 0.87 when the dynamic traces are generated with humans stimulating the apps compared to 0.86 and 0.85 achieved respectively by static and dynamic (humans) analysis alone. In {\tt package} mode, the hybrid system achieves $F$-measure of 0.92 when %
the dynamic traces are produced %
by Monkey %
and 0.90 %
with humans. %
The hybrid system outperforms the dynamic analysis system in all modes (i.e., {\tt family} and {\tt package}), as it also captures behavior not exhibited during runtime execution of the apps as a result of the overestimation from static analysis, while it improves static analysis as it captures frequently used API calls -- a behavior that cannot be captured by static analysis -- and API calls that are dynamically loaded.

\subsection{Misclassifications within each analysis method}

\begin{figure}[t!]
\centering
\subfigure[\label{fig:falsepositives} Humans] 
{\includegraphics[width=0.215\textwidth]{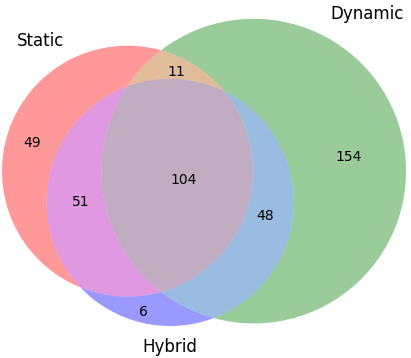}}
\subfigure[\label{fig:falsepositives1} Monkey] 
{\includegraphics[width=0.215\textwidth]{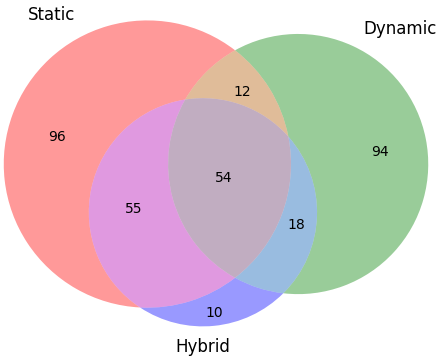}}
\vspace{-0.3cm}
\caption{Number of false positives in each analysis method and when the apps are stimulated by humans (a) or Monkey (b) during dynamic analysis.}
\label{fig:miss_set}
\vspace{-0.3cm}
\end{figure}

Next, we examine the samples that are misclassified 
in each method of analysis, aiming to understand the
differences in the model 
of the correctly classified and misclassified samples.
We perform our analysis on samples that have been classified by all three methods 
in {\tt package} mode. 

We formulate and verify the %
hypothesis that misclassifications are due to missing API calls that are considered ``important'' by the classifiers. To this end, %
we select the 100 most important features used by each classifier to distinguish between potential malware and benign samples, and evaluate the average number of these features present in each sample. %
We select the 100 most important features because it %
represents, at most, about 10\% of the features %
recorded in our experiments. Recall that a feature in our detection technique is the probability of evoking an abstracted API call, and transitions not evoked during the experiments have probability of 0. The maximum number of features with probability $>$ 0 in our dataset is 1,869 (static analysis) and the minimum is 1,022 (dynamic analysis with humans). We expect that samples that are misclassified will have a similar number of important features as those of the opposite class. 
Therefore, using the top 100 features for each classifier, we compare the average number of the features in the true positives (i.e., correctly classified malware samples) %
to the false negatives (malware classified as benign), as well as true negatives to false positives. 

\begin{figure*}[t!]
\centering
\subfigure[\label{fig:staticH} Static Analysis]{\includegraphics[width=0.28\textwidth]{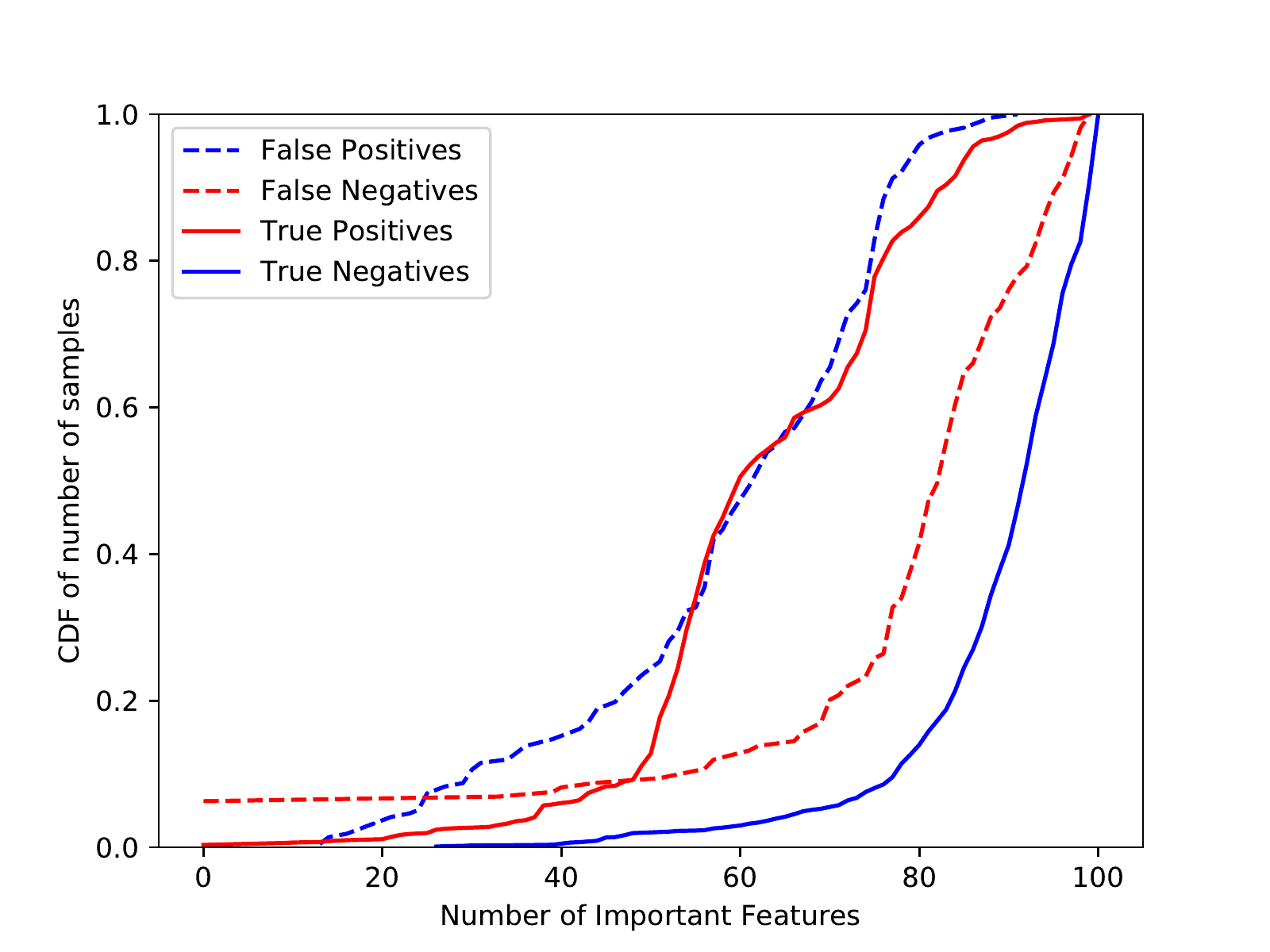}}
\subfigure[\label{fig:dynamicH} Dynamic Analysis]{\includegraphics[width=0.28\textwidth]{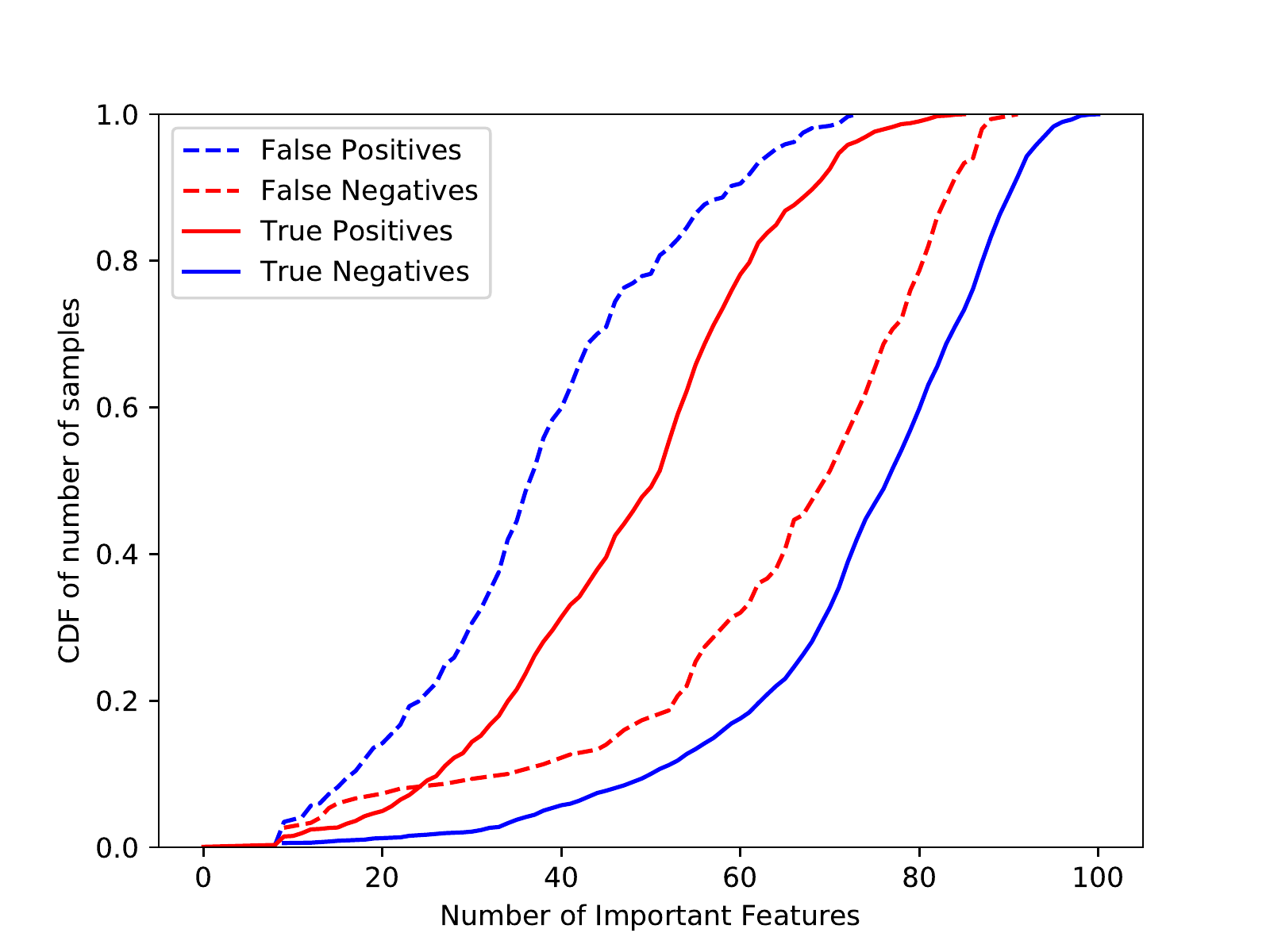}}
\subfigure[\label{fig:hybridH} Hybrid Analysis]{\includegraphics[width=0.28\textwidth]{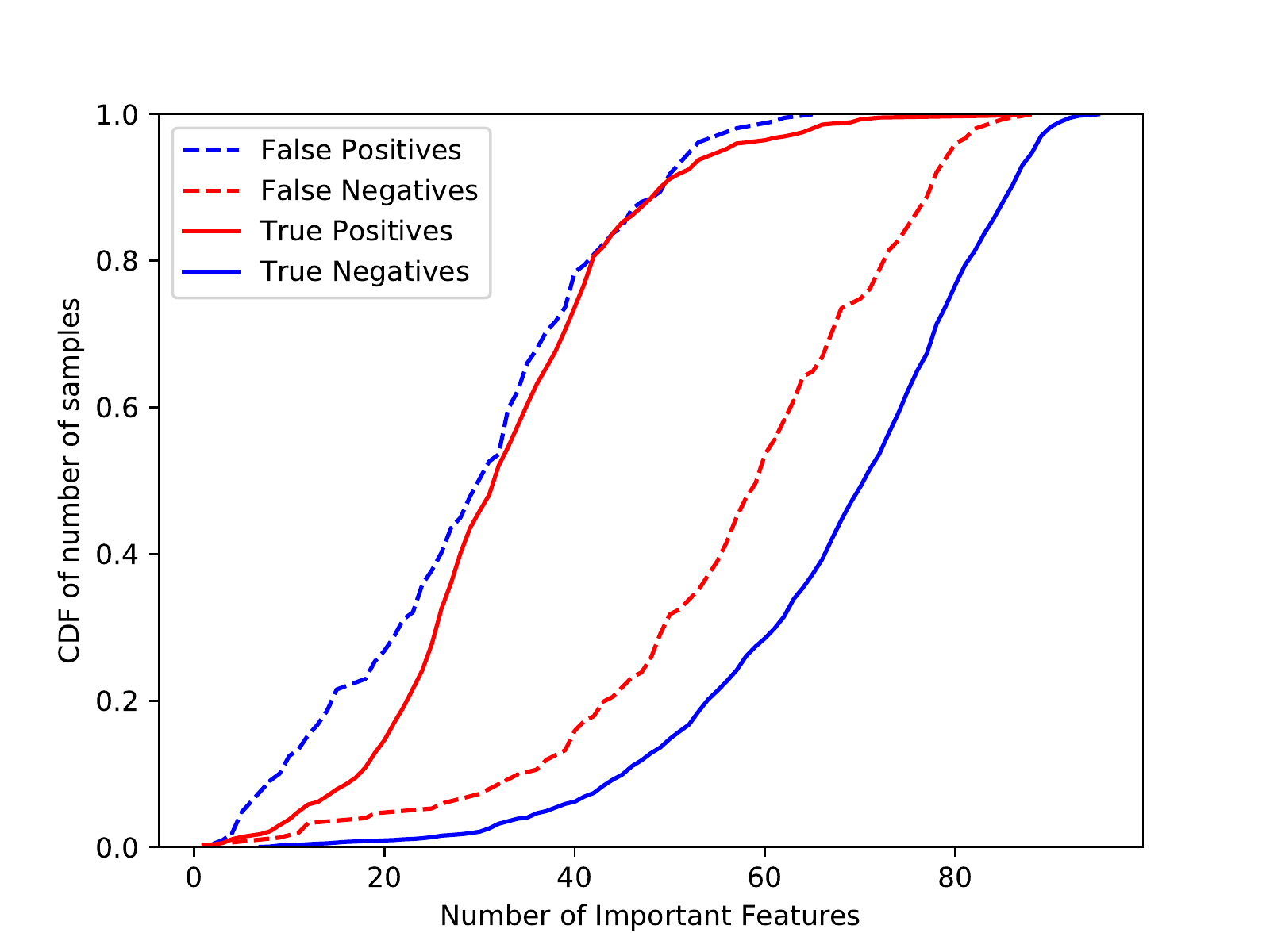}}
\caption{CDF of the number of features present (out of the 100 most important features) in each classification type for all analysis methods, with {\bf human}, during dynamic analysis.}
\label{fig:topFeaturesH}
\end{figure*}

\begin{figure*}[t!]
\centering
\subfigure[\label{fig:staticM} Static Analysis]{\includegraphics[width=0.28\textwidth]{figs/staticAcc.pdf}}
\subfigure[\label{fig:dynamicM} Dynamic Analysis]{\includegraphics[width=0.28\textwidth]{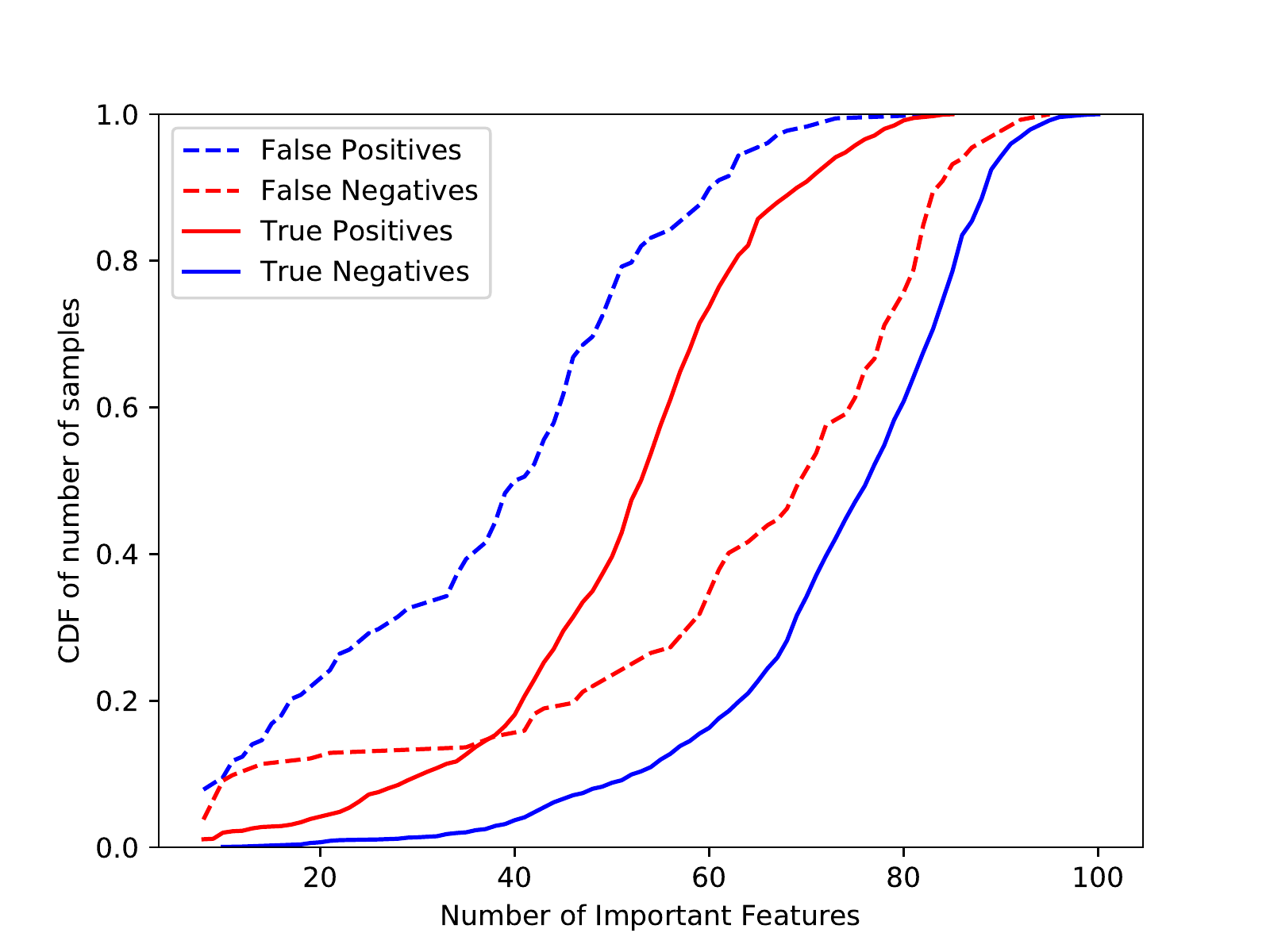}}
\subfigure[\label{fig:hybridM} Hybrid Analysis]{\includegraphics[width=0.28\textwidth]{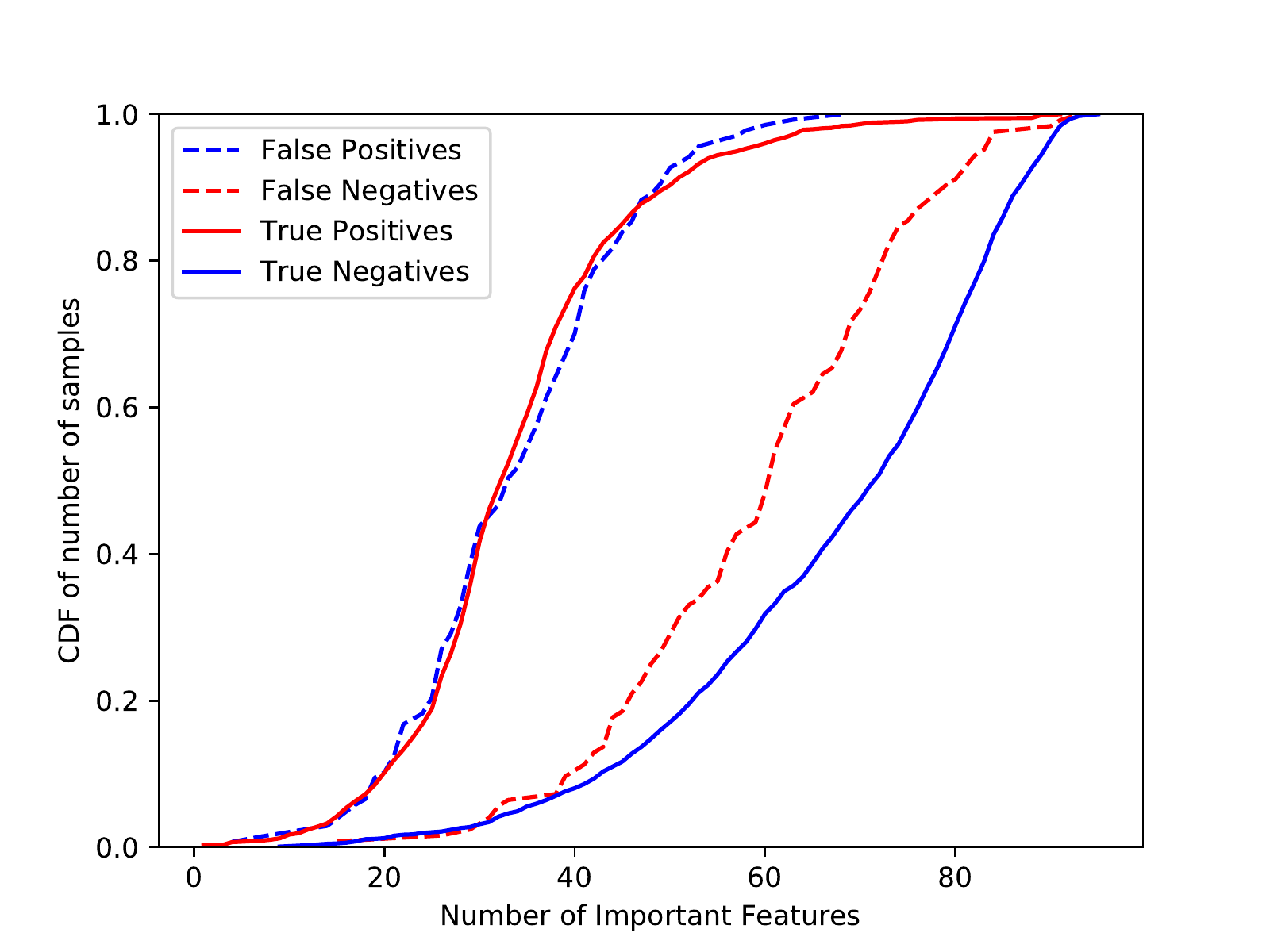}}
\caption{CDF of the number of features present (out of the 100 most important features) in each classification type for all analysis methods, with {\bf Monkey}, during dynamic analysis).}
\label{fig:topFeaturesM}
\end{figure*}

\descr{False Positives.} 
In \figurename~\ref{fig:falsepositives} and~\ref{fig:falsepositives1}, we report 
the number of false positives (i.e., benign samples classified as malware) in each method of analysis, respectively, when apps are stimulated by humans and by Monkey during dynamic analysis. With the former, there are 215, 317, and 209 false positives, respectively, with static, dynamic, and hybrid analysis. With the latter, we get 217, 178, and 137 false positives with static, dynamic, and hybrid analysis.
Using the top 100 features, we find that the false positives %
in static analysis exhibit similar behavior to that observed in true positives. Specifically, they
have, on average, 54.12$\pm$22.65 features out of the 100 most important features, which is similar to %
59.96$\pm$19.46 in true positive samples. %
The same behavior is also observed in both dynamic and hybrid analysis irrespective of the app stimulator. 
In \figurename~\ref{fig:topFeaturesH}, we plot the CDF of the number of features present in each classification type for all analysis methods when humans %
stimulate apps during dynamic analysis and, in \figurename~\ref{fig:topFeaturesM}, %
with Monkey. %
That is, the behavioral model of the false positives in all analysis methods is similar to that observed on the true positives. For example, in \figurename~\ref{fig:hybridH} (hybrid analysis) 90\% of the false positives have no more than 50 of the 100 most important features (similar to the true positives -- 49/100) while true negatives reach 86 features out of 100.

\begin{figure}[t!]
\centering
\subfigure[\label{fig:falsenegatives} Humans] 
{\includegraphics[width=0.215\textwidth]{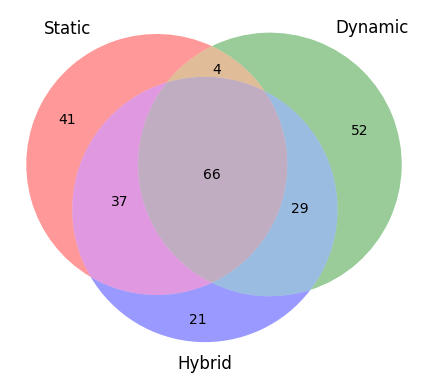}}
\subfigure[\label{fig:falsenegatives1} Monkey] 
{\includegraphics[width=0.215\textwidth]{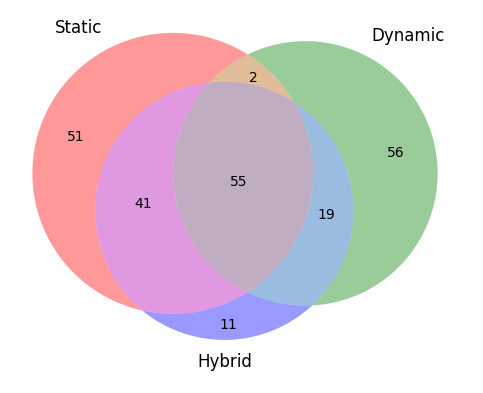}}
\caption{Number of false negatives in each analysis method and when the apps are stimulated by humans (a) or Monkey (b) during dynamic analysis.}
\label{fig:miss_set1}
\end{figure}

\descr{False Negatives.} 
In \figurename~\ref{fig:falsenegatives} and~\ref{fig:falsenegatives1}, we report 
the number of false negatives (i.e., malware samples classified as benign) in each analysis method, resp., when apps are stimulated by humans and Monkey. With the former, there are 148, 151, and 153 false negatives, respectively, in static, dynamic, and hybrid analysis, while, with the latter, we get 149, 132, and 126 false negatives.
In static analysis, we find that the behavioral model of the false negatives are similar to that observed in the true negatives. In particular, of the 100 most important features used to distinguish malware from benign samples, there are, on average, 82.08$\pm$11.75 features per false negative sample. The value is more similar to the 88.91$\pm$11.31 important features per true negative sample rather than the 59.96$\pm$19.46 important features per true positive sample. The same result is also observed in dynamic analysis irrespective of the stimulator, and in hybrid analysis as well. Recall that, in \figurename~\ref{fig:topFeaturesH} and~\ref{fig:topFeaturesM}, we plot the CDF of the number of features in each classification type when, resp., human and Monkey are used as the stimulator during dynamic analysis; e.g., in \figurename~\ref{fig:dynamicM} (dynamic analysis), 90\% of the false negative samples have 84 of the 100 features, a value more similar to 89 features (true negatives) rather than 70 features (true positives).

\subsection{Misclassifications across analysis methods}
Next, we attempt to clarify why some samples are misclassified by one method of analysis but correctly classified by another. 
The first important difference among the methods is the code coverage: dynamic analysis does not cover the entire code base of an app. Moreover, stimulating with Monkey %
vs %
humans yield different code coverage. This might result in a few different scenarios: 
\shortVer{1) Dynamic analysis may not have triggered the malicious code that  is captured in static analysis;
2) Static analysis may reveal sequences of API calls that are not necessarily malicious, but characterize many malicious apps;
3) API calls not triggered during dynamic analysis may affect the Markov chains leading to training poisoning or misclassification of the sample, depending on whether the sample is part of the training or the test set.}
\longVer{
\begin{enumerate}
\item Dynamic analysis may not have triggered the malicious code that  is captured in static analysis;
\item Static analysis may reveal sequences of API calls that are not necessarily malicious, but characterize many malicious apps;
\item API calls not triggered during dynamic analysis may affect the Markov chains leading to training poisoning or misclassification of the sample, depending on whether the sample is part of the training or the test set.
\end{enumerate}}

Scenarios (1) and (2) are possible reasons why static analysis correctly detects some samples and dynamic analysis does not, while  (3) refers to the opposite. 
Although the hybrid system captures sequences of API calls from both static and dynamic 
analysis, %
it actually results in completely new Markov chains and features for training and classification. %
While more accurate than the individual methods, as the features values change (i.e., the transition probabilities), it behaves differently.

Another important factor is the presence of loops in the code waiting for user interaction. A good example are the
games \textit{Dumb ways to die 1 \& 2}. These apps have %
``minigames'' where a user has to click several times on the right spot of the screen at the right time. %
When executed, the apps enter a loop waiting for user action, and decide the next action based on what happened before returning to waiting for user action. Static analysis would catch the four different outcomes (i.e., execution path) of the loop, i.e., wrong click, correct click, the user won the game, the user lost the game. Dynamic analysis would repeat the loop many times depending on the continuous clicks of the human or of Monkey, and the user/Monkey may never win or lose.
Static analysis will record the four possible loop paths without repeating the sequences in its traces,
and the user/Monkey may not record all the possible sequences, but 
have duplicated sequences due to multiple clicks resulting in the same outcome. 
All these differences characterize the recorded traces, and therefore may result in different Markov chains and decisions among the methods. %

\section{Discussion \& Conclusion}
\label{sec:conclusion}
In this paper, we analyzed different Android malware detection analysis methods, i.e., static, dynamic, and hybrid analysis,
using a common modeling approach. Specifically, we built a behavioral model of each sample based on the sequences of abstracted API calls, as done by \mama~\cite{mariconti2016mamadroid}, as it effectively captures malicious behavior even in the presence of changes in the Android API and evolving malware. %
We then introduced a dynamic analysis tool, \papa, which supports app stimulation via both humans (via crowdsourcing~\cite{mario2018chimp}) and pseudorandom input generators (Monkey). We also slightly modified \mama to first abstract an API call to its class, before abstracting to other modes, to avoid abstracting to the wrong package.
Then, to build a hybrid system, we merged the sequences of API calls from static and dynamic analysis. 
All three methods operate in one of two modes, i.e., {\tt family} and {\tt package}, based on the level of abstraction; 
in {\tt family} mode, static, dynamic (human/Monkey), and hybrid analysis, respectively, achieve $F$-measures of 0.86, 0.85/0.86, and 0.88. Whereas, in {\tt package} mode, we achieve 0.91, 0.88/0.92, and 0.92.

Overall, our experiments showed that hybrid analysis performs best because it captures the best of static and dynamic analysis,
as it is able to capture the sequences of API calls that are actually executed and/or dynamically loaded (from the latter), and capture code not executed during testing due to code overestimation (from the former). 
Nonetheless, static analysis performs well overall, often better than dynamic analysis;
when looking at misclassifications across methods, we found that those occurring in dynamic but not in static analysis are likely due to poor code coverage, thus, the feature vectors in dynamic analysis may not reveal features (e.g., a chunk of benign code in repackaged samples) that characterize malware in our dataset.
Finally, we showed that dynamic analysis performs better with Monkey than humans because the former is able to trigger more code than the latter.

Although some characteristics peculiar to \papa's virtual device (e.g., %
it runs as a hardware assisted virtualization) should prevent evasion by malware that tries to circumvent emulators/virtual devices using environment variables~\cite{jing2014morpheus,Maier2014,diao2016evading}, %
we plan, as part of future work, to update it to use a virtual device that appears as close to a real device as possible. 
In Appendix~\ref{limitations}, we also highlight some of its limitations, which we will address in the near future.
Moreover, we intend to use input generators that target specific behaviors of an app, so as to target certain API calls mostly used by malware rather than trying to improve the code coverage during dynamic analysis. %
Finally, we plan to detect and measure the prevalence of malware that specifically employs dynamic code loading as an evasion technique. %

\descr{Acknowledgments.} Lucky Onwuzurike was funded by the Petroleum Technology Development Fund (PTDF), Enrico Mariconti was supported by the EPSRC under grant 1490017.

{\small
\bibliographystyle{abbrv}

}

\appendix

\section{Challenges With Dynamic Analysis}
\label{challenges}
We now discuss the challenges we faced when analyzing the apps during dynamic analysis.

\descr{Apps with no trace.}
\label{failures}
When executing apps during dynamic analysis %
on \papa, we find that, for some apps, we do 
not obtain any trace. Specifically, %
there are no traces for 724 apps when using Monkey %
and 693 when using humans. In total, we do not obtain traces for 835 unique apps.
Further examination of the logs of these apps shows that this happens because:
\begin{enumerate}
\item The apps stop responding, i.e., ``Application Not Responding'' (ANR) is thrown by Android as the app cannot respond to user input (154 occurrences); %
\item Lack of OpenGL minimum specification matching the requirement of the app (229);
\item Fatal error in an app's {\tt onCreate()} launcher activity method, causing the app to crash (452).
\end{enumerate}

\descr{Apps with no package name in apk.} %
For some apps where we do obtain traces, we find that they do not contain any API call beginning with the package name of the app. Recall that in order to evaluate the code coverage of each app, we focus on API calls that begin with the package name of the app. %
Using this approach, we find there are 350 apps in our dataset for which calls beginning with the package name are not present in the dynamic traces nor the apk. Upon further analysis, we find these apps use package names that are different from that declared in their manifest. To calculate code coverage for these apps, we use as package names those packages (excluding packages from the Android and Google API) that have at least one activity and broadcast receiver and/or service classes in the apk.

\section{Virtual Device Limitations}
\label{limitations}

One of the contributions of our work is  %
extending \chimp~\cite{mario2018chimp} to create a virtual device on which \papa builds;
naturally, our work is not without limitations.
First, we only support one Android version at this time: KitKat 4.4.2,
however, we expect KitKat to actually be more vulnerable to malware than more recent versions of Android; moreover, only 0.8\% of apps in our sample  (see Section~\ref{sec:preprocessing}) require a newer version of Android.
An interesting future avenue would be to explore differences in malware behavior across versions of Android, which previous work showed to be quite significant~\cite{xing2014upgrading}.

Second, our virtual device does not yet support the full range of OpenGL operations, which somewhat limits the number of apps that we can evaluate. However, this support is an ongoing effort by the Android x86 community\footnote{Qemu's OpenGL 3 support was announced very recently -- see \url{http://www.android-x86.org/releases/releasenote-7-1-rc1}} and only 229 (4.6\% of our dataset) of apps that we failed to acquire a trace for required full OpenGL support.
Finally, we remark that, when stimulating the apps using humans, the users we recruit may not have tested each app in the way they would have used the apps on their own device, which is also an item for future work.

\end{document}